\documentclass[fleqn,usenatbib]{mnras}

\usepackage{newtxtext,newtxmath}

\usepackage[T1]{fontenc}

\DeclareRobustCommand{\VAN}[3]{#2}
\let\VANthebibliography\thebibliography
\def\thebibliography{\DeclareRobustCommand{\VAN}[3]{##3}\VANthebibliography}

\usepackage{graphicx}	
\usepackage{amsmath}	
\usepackage{float}
\usepackage[dvipsnames]{xcolor}

\newcommand{\kms}{\ensuremath{\, \mathrm{km}\, {\mathrm s}^{-1}}}

\newcommand{\ergs}{\ensuremath{\, \mathrm{erg}\, {\mathrm s}^{-1}}}

\title[Stellar expansion for GW sources]{The role of stellar expansion on the formation of gravitational wave sources}

\author[A. Romagnolo et al.]{
A. Romagnolo$^{1}$,\thanks{E-mail: amedeoromagnolo@gmail.com}
K. Belczynski$^{1}$,\thanks{E-mail: chrisbelczynski@gmail.com}
J. Klencki$^{2,3}$,
P. Agrawal$^{4}$,   
T. Shenar$^{5}$,
D. Sz\'ecsi$^{6}$
\\
$^{1}$Nicolaus Copernicus Astronomical Center, The Polish Academy of Sciences, ul. Bartycka 18, 00-716 Warsaw, Poland\\
$^{2}$European Southern Observatory, Karl-Schwarzschild-Strasse 2, 85748, Garching bei München, Germany\\
$^{3}$Max Planck Institute for Astrophysics, Karl-Schwarzschild-Strasse 1, 85748, Garching bei München, Germany\\
$^{4}$ McWilliams Center for Cosmology, Department of Physics, Carnegie Mellon University, Pittsburgh, PA 15213, USA\\
$^{5}$Anton Pannekoek Institute for Astronomy, Science Park 904, 1098 XH, Amsterdam, the Netherlands\\
$^{6}$Institute of Astronomy, Faculty of Physics, Astronomy and Informatics, Nicolaus Copernicus University, Grudzi\k{a}dzka 5, 87-100 Toruń, Poland
}

\date{Accepted XXX. Received YYY; in original form ZZZ}

\pubyear{2022}

\begin{document}
\label{firstpage}
\pagerange{\pageref{firstpage}--\pageref{lastpage}}
\maketitle

\begin{abstract}
Massive stars are the progenitors of black holes and neutron stars, the mergers of which can be detected with gravitational waves (GW). The expansion of massive stars is one of the key factors affecting their evolution in close binary systems, but it remains subject to large uncertainties in stellar astrophysics. 
For population studies and predictions of GW sources, the stellar expansion is often simulated with the analytic formulae from Hurley et al. (2000). These formulae need to be extrapolated and are often considered outdated. 

In this work we present five different prescriptions developed from 1D stellar models to constrain the maximum expansion of massive stars. We adopt these prescriptions to investigate how stellar expansion affects mass transfer interactions and in turn the formation of GW sources.

We show that limiting radial expansion with updated 1D stellar models, when compared to the use of Hurley et al. (2000) radial expansion formulae, does not significantly affect GW source properties (rates and masses). This is because most mass transfer events leading to GW sources are initialised in our models before the donor star reaches its maximum expansion. The only significant difference was found for the mass distribution of massive binary black hole mergers ($M_{\rm tot}$ > 50\,M$_\odot$) formed from stars that may evolve beyond the Humphreys-Davidson limit, whose radial expansion is the most uncertain. We conclude that understanding the expansion of massive stars and the origin of the Humphrey-Davidson limit is a key factor for the study of GW sources.


\end{abstract}




\section{Introduction}
\label{sec:intro}

The LIGO/Virgo/KAGRA collaboration (LVK) has so far detected around 76 binary black hole (BH-BH), 12 binary neutron star (NS-NS) and 2 black hole-neutron star (BH-NS) mergers in its three observing runs \citep{Abbott2021}. The main observables that can be derived from these mergers are the merger rate density in the local universe (redshift $z\sim$ 0.2)  and the distribution of masses and orbital spins. The reported 90\% credible intervals of the merger rate densities from LVK observations are 16-61 Gpc$^{-3}$yr$^{-1}$ for BH-BH, 7.8-140 Gpc$^{-3}$yr$^{-1}$ for BH-NS and 10-1700 Gpc$^{-3}$yr$^{-1}$ for NS-NS. In the case of NSs there is no pronounced single peak in the mass distribution and their inferred maximum mass goes up to $\sim$ 2.2\,M$_\odot$ within the margins of uncertainty. In contrast, the distribution of the BH primary masses (the masses of the more massive BH in the binary) was shown to have substructures beyond an unbroken single power law, where two local peaks at $\sim$ 10 and 35\,M$_\odot$ are found. Finally, the effective spin distribution has been inferred in \cite{Abbott2021} to have a mean centred at $\sim$ 0.06.\\
\\
There are several evolutionary channels that have been proposed to date to explain LVK observations. Recent studies suggest that most of massive stars will interact with a companion star in their lives \citep{Sana2012,Moe2017}. A predominant channel is therefore isolated binary evolution, where populations of massive binaries that evolve in relative isolation are considered. There are in turn several sub-channels for isolated binary evolution that could lead to close double compact objects. One of the most studied ones is the Roche lobe overflow (RLOF) scenario, where two massive stars born in a wide orbit have sufficient space to evolve. In this configuration one of the main ingredients dictating the evolution of the two stars, their collapse into compact objects (COs), and their potential merger is their radial evolution. If during the evolution one of the component stars becomes larger than its Roche lobe radius it will initialise a RLOF event. These events lead to a redistribution of the binary total mass and also to its partial expulsion, which translate into a loss of angular momentum for the binary system. These two factors in turn will decrease the orbital distance and potentially initialise other RLOF events \citep{Paczynski_1976,Belczynski_2002,Olejak_2021,mandel2021rates}. On top of that, if a hydrogen layer is still present, low-metallicity stars that were stripped in binaries can still expand to giant sizes despite having lost most of their envelopes, and therefore potentially re-initiating a RLOF event \citep{Laplace_2020}.\\
\\
Alternatively, in case of an unstable RLOF due to a combination of mass ratio between the two binary components, the presence of a convective envelope in the donor star, metallicity ($Z$) and stellar type \citep{Pavlovskii_2017,Klencki_2020,Klencki_2021,Olejak_2021} the binary will face a common envelope (CE) phase. In this case friction between the two stellar objects and the envelope will further reduce the orbital momentum of the system. This is extremely important as the closer the stars are, the more likely is that they will merge within a Hubble time after their collapse into compact objects. 
Another alternative is the chemically homogeneous evolution sub-channel, where massive binary stars in a close orbit do not significantly expand due to the influence of efficient internal mixing. In this scenario massive stars in their main sequence (MS) mix the helium of their core out into the envelope and bring hydrogen from the envelope into the core. This will bring the star to burn almost all its hydrogen and contract at the end of its MS and not significantly expand throughout the rest of its life \citep{Heger_2000,MaederMeynet2000}. Systems evolving this way do not face any RLOF event and create heavy BH binaries at low orbital separations that can merge within a Hubble time \citep{MandeldeMink_2016,Marchant_2016,deMinkMandel_2016}, but they are quite rare \citep{mandel2021rates}.\\
\\The dynamical formation channel treats stars and binaries in dense stellar environments where gravitational interactions with other stars influences their evolution. Sub-channels for the dynamical formation are usually divided based on the considered environment. In the young star cluster and the globular cluster sub-channels, single BHs sink towards the centre of the cluster through mass segregation \citep{Spitzer_1969} and either form a binary with another BH or influence the evolution of already existing compact binaries. Nuclear star clusters in the centre of galaxies, in the absence of supermassive black holes, have higher escape velocities than globular clusters \citep{Miller_2009,Antonini_2016} and therefore are a promising nursery for the birth of BH-BH binaries. On top of that the influence of three-body scattering and gaseous drag in active galactic nuclei could deeply influence the birth and merger rates of BH-BH binaries \citep{Bellovary_2016,Stone_2016,Bartos_2017,McKernan_2018,Tagawa_2020}. The presence of a supermassive black hole, instead, would make the escape velocity of the cluster so high to counter recoil kicks from BH-BH mergers and therefore to favour hierarchical mergers \citep{Yang_2019,Secunda_2020,Tagawa_2021}.\\
\\Another proposed dynamical formation scenario involves hierarchical triple systems where, according to the Kozai-Lidov mechanism \citep{Kozai_1962,Lidov_1962}, the exchange of angular momentum between the inner binary and the orbit of the outer companion periodically alters the eccentricity of the inner binary, which could catalyse the merging process (e.g. \citealt{Liu_2018}).\\ 
\\Finally, a potential GW source channel involves primordial BH binaries, which are BHs of any mass beyond the Planck mass that were born from energy fluctuations during the early stages of the Universe, and that are mass-wise potentially enhanced by dark matter \citep{Bird2016,Haimoud2017,Chen_2018}.\\
\\
There have been already some efforts to develop theoretical environments and specific computational methods to infer population models that could combine multiple formation channels \citep{Ng_2021,Wong_2021,Zevin_2021}. An important point to highlight is that at the current stage we cannot draw any conclusion regarding the respective contribution of each evolutionary channel for the observed LVK merger rates, since the considerable sources of uncertainty in each model hinder the development of a clear picture to explain the observational data \citep{Belczynski2022}. \cite{Mandel_2022} showed that not only between different evolutionary channels, but also in the same evolutionary channel with different model implementations and codes, the estimates for CO mergers can vary of several orders of magnitude.\\
\\
Stellar evolution influences stellar radius and, in turn, is influenced by it.
One of the most important factors that determine the radial evolution of a star of a given mass is metallicity \citep{Xin_2022}. For instance, the convective eddies extension, as described by the Mixing-Length Theory (MLT), was shown to be dependent on metallicity \citep{Bonaca_2012}, which therefore means that the efficiency of stellar internal mixing is as well dependent on it. Low metallicity levels affect the opacity and the nuclear burning processes inside the star. The lower the opacity, the more easily the luminosity produced in the inner regions of a star will be able to escape. Low opacity also means low radiative pressure, which means that the outer layers reach hydrostatic equilibrium deeper in the star where the temperature and pressure are higher \citep{Burrows_1993,Kesseli_2019}, but a low metallicity also means a suppression of strong winds. Internal mixing also influences stellar radius evolution, since the transport of different elements affects both the mean molecular weight and the radiative opacity of the stellar envelope. In general enhanced internal mixing during the main sequence reduces the mass threshold for a star to remove its envelope through winds \citep{Gilkis_2021}, thereby preventing a considerable stellar expansion. The expansion of stars after the end of main sequence has been shown to be even more uncertain due to it being sensitive to the (poorly constrained) composition profile in the envelope layers above the H-burning shell\citep{Georgy_2013,Klencki_2020,farrell2021}.\\
\\
A key problem in our understanding of the evolution of massive stars is their behaviour beyond the so-called Humphrey\,-\,Davidson (HD) limit. This limit consists of a threshold in the Hertzsprung-Russel (HR) diagram beyond which no (or very few) stars have been observed, despite what many evolutionary models may predict \citep{Humphreys_1994} and it so far apprears to be independent on metallicity \citep{Davies2018,Gilkis_2021} The existence of the HD limit might be explained with the surface of massive stars being in close proximity to their respective Eddington limit. Under this condition, the surface regions are subject to  hydrostatic unbalance, turbulence and large mass ejection events up to 10$^{-3}$\, M$_\odot$\,yr$^{-1}$ \citep{Owocki_2004,Agrawal_2020}. This behaviour has been linked to an observed class of stars, namely luminous blue variables (LBV), which experience episodes of large mass loss events \citep{Humphreys_1994,Schaerer_1995,Smith_2004}.\\
\\
Many modern population synthesis models are based on the analytic formulae from \cite{Hurley_2000}, which describe the evolution of stars as a function of mass and stellar metallicity. These formulae were fitted from a set of models with initial stellar masses below 50\,M$_\odot$. Simulating the progenitors of very massive BHs in this framework requires therefore extrapolation, which could lead to potential artefacts. Given all the relatively new observational and theoretical data since the publication of these formulae, population synthesis models need to implement all the information available from survey missions and detailed evolutionary codes in order to further constrain stellar evolution and in turn to correct for any potential artefact that may arise from the use of these models for physical conditions they weren't developed for. For example, \cite{Laplace_2020} showed that population synthesis codes (and also some sets of detailed stellar models) fail to account for the structure of stripped stars, which leads to underestimations of the rate of late mass transfer events up to an order of magnitude.
Recent studies have shown how the radial evolution of massive stars is highly uncertain, to a point that, for a given Zero-Age Main Sequence (ZAMS) mass $M_{\text{ZAMS}}$ and metallicity \textit{Z}, two different models could simulate stars that differ by even more than 1000\,R$_\odot$ \citep{Agrawal_2020,Agrawal:2022,Belczynski2022}. The first population synthesis study that tackled the role of stellar expansion during the giant phase for binary evolution was \cite{FWH1999}, which analysed the uncertainties that could lead to different NS-NS merger rates and showed how these rates can drastically vary as a function of stellar radii and kick velocities. {\cite{Mennekens_2014} also tackled this topic by studying how the wind-driven mass loss during the LBV phase could deplete the H-rich envelope for post-MS stars with $M_{\rm ZAMS}\gtrsim$\,40\,M$_\odot$. They then analysed how this depletion could stop massive stars from evolving into red supergiants and in turn how this can affect the formation of close BH-BH binaries and their mergers.\\
\\
In our study we examine the radial expansion of massive stars in the context of GW progenitors from isolated binary evolution. We analyse how the maximum stellar radius changes as a function of metallicity and ZAMS mass according to \cite{Hurley_2000} formulae and according to one dimensional (1D or detailed) stellar evolution codes.
We then develop four new prescriptions by fitting evolutionary equations from detailed evolutionary simulations in order to simulate the maximum stellar expansion of massive stars ($M_{\text{ZAMS}}$ > 18\,M$_\odot$) as a function of their ZAMS masses. We then adopt these new prescriptions in our population synthesis pipeline to study the impact of different radial expansion models on the estimates for double compact object merger rate densities and BH-BH mass distribution for LVK observations.


\section{Method}
\label{sec:Method}

In the following section we describe five different codes and calculations (either performed in this study or adopted from literature) that we use to show how different prescriptions can alter the maximum radii of massive stars and the estimates of double compact object merger rate densities and BH-BH merger mass distributions.\\
\\
In our study we use the {\tt StarTrack}\footnote{\url{https://startrackworks.camk.edu.pl}} \citep{Belczynski_2002,Belczynski_2008} population synthesis code to simulate the formation and mergers of double compact objects. It allows us to predict how a population of binary stars behave for a wide array of initial conditions and physical assumptions, but instead of re-computing their evolution from first principles on the fly it relies on the evolutionary formulae from  \cite{Hurley_2000}. The currently implemented physics, star formation history, Universe metallicity and its evolution with cosmic time are described in \cite{Belczynski_2020}, with two modifications described in Sec. 2 of \cite{Olejak_2020}.\\
\\
To constrain the maximum stellar radius ($R_{\rm MAX}$) within {\tt StarTrack} simulations we derive analytic formulae from four different sets of simulations: (i) the \textsc{`Bonn’ Optimized Stellar Tracks} ({\tt BoOST}) from  \citet{szecsi2021} (ii) the \textsc{Modules for Experiments in Stellar Astrophysics} \citep[MESA;][]{Paxton2011, Paxton2013, Paxton2015, Paxton2018, Paxton2019} simulations from \citet{Agrawal_2020} and (iii) two different setups for our {\tt MESA} simulations. 
Both BoOST and MESA simulations from \citet{Agrawal_2020} were further interpolated by the \textsc{Method of Interpolation for Single Star Evolution} \citep[METISSE;][]{Agrawal_2020} to create a smooth distribution of radius with initial stellar masses. These prescriptions are meant to limit the maximum radius that a star can reach throughout its whole lifetime as a function of its ZAMS mass. It must be stressed that our prescriptions do not alter the simulated stellar expansion in any way. We simply set a hard limit to how much a star could grow in size as a function of its ZAMS mass.\\
\\
Strong stellar winds at high metallicities (see e.g. \citealt{Heger_2003}) and mass exchange during RLOF events from the donor star to its companion can alter whether a star will evolve into a NS or a BH. \cite{Muller_2016} also showed that for a given metallicity there might not be a hard maximum $M_{\rm ZAMS}$ value beyond which a massive star will always collapse into a BH. We nevertheless choose to fit and apply these $R_{\rm MAX}$ formulae for a $M_{\rm ZAMS}$ range between 18 and 150 M$_\odot$ i) in order to affect only BH and massive NS and ii) because stellar evolution is less understood at these masses.

\subsection{Model 1 -- StarTrack}
We adopted the broken power-law for the initial mass function from \cite{Kroupa_1993} \& \cite{Kroupa_2002} for primary stars\footnote{We define as a primary star the more massive binary component at ZAMS.} with the following exponents:\\
$\alpha_1$ = -1.3 for $M_{\rm ZAMS}$ $\in$ [0.08; 0.5]\,M$_\odot$,\\
$\alpha_2$ = -2.2 for $M_{\rm ZAMS}$ $\in$ [0.5; 1]\,M$_\odot$,\\
$\alpha_3$ = -2.3 for $M_{\rm ZAMS}$ $\in$ [1; 150]\,M$_\odot$\\
The mass of the secondary star is the mass of the primary companion (M1) multiplied by a mass ratio factor q from a uniform distribution \textit{q} $\in$ [0.08/M1; 1] (for mass distributions see e.g. \citealt{Shenar_2022}). To produce the semi-major axis of the binary system we use the initial orbital period (P) power law distribution log(P [days]) $\in$~[0.5;~5.5] with an exponent $\alpha_P$ = -0.55 and a power law initial eccentricity distribution with an exponent $\alpha_e$ = -0.42 within the range [0; 0.9], and invoke Kepler's third law. The initial orbital parameters are taken from \cite{Sana2012}, but we adopt the extrapolation of the orbital period distribution from \cite{Mink_2015}, where the period range has been extended to log(P) = 5.5 . The stellar winds prescription adopted in \textsc{StarTrack} is based on the theoretical predictions of radiation driven mass loss from \cite{Vink2001}. So far no self-consistent model has been developed to prescribe in detail the LBV mass loss. Our prescription for the LBV mass loss $\dot{M}_{\rm lbv}$ is adopted from \cite{Belczynski_2010b}: 

\begin{equation}
    \dot{M}_{\rm lbv}=f_{\text{lbv}} 10^{-4}[{\rm M}_\odot {\rm yr}^{-1}]  \\\\f_{\rm lbv}=1.5
    \label{eq:lbv}
\end{equation}

This formula is meant to be an average mass loss that accounts for both LBV stellar wind mass loss and possible LBV shell ejections. The LBV factor $f_{\rm lbv}$ was calibrated in \cite{Belczynski_2010b} so that it could reproduce the most massive Galactic BHs that were known at that time (e.g. \citealt{Orosz_2003,Casares_2007,Ziolkowski_2008}). We stress that this calibration was made prior to the first GW detection and to the most recent BH surveys from electromagnetic observations, which implies that $f_{\rm lbv}$ might need to be re-calibrated to be up to date. In order to define under which conditions our LBV mass loss prescription from Equation \ref{eq:lbv} must be adopted, we use the \cite{Hurley_2000} definition for the Humphrey-Davidson limit:
\begin{equation}
    \begin{cases}
      {\rm log}L > 6\times10^5\\
      10^{-5}\times R \times \sqrt{L} > 1 \hspace{0.2 cm} R_\odot L^{1/2}_\odot
    \end{cases}
\label{eq:HDlimit}
\end{equation}

With log as a logarithm base 10, $L$ stellar luminosity in L$_\odot$ and $R$ stellar radius in R$_\odot$.}To calculate the mass accretion from stellar winds we use the approximation from \cite{BoffinJorissen1988} of the \cite{BondiHoyle1944} model (for more information see \citealt{Belczynski_2008}). The adopted prescription for the accretion onto a CO both during a stable RLOF event or from stellar winds is based on the approximations from \cite{King2001}. We use 50 per cent non-conservative RLOF for non-degenerate accretors, with a fraction of accreted donor mass $f_{\rm a}$ = 0.5 and the remaining mass (1 $-f_{\rm a}$) being dispersed in the binary surroundings with a consequent loss of orbital angular momentum \citep{Belczynski_2008}. In order to determine the potential instability of a RLOF event (and therefore a CE phase) we adopt the diagnostic diagram described in \cite{Belczynski_2008}.
To treat CE events we use the Alpha-lambda prescription from \cite{Webbink_1984}, with $\alpha_{CE}=1$. We obtain the $\lambda$ value in {\tt StarTrack} from the so-called Nanjing $\lambda$ procedure \citep{Dominik_2012}, which comes from the \cite{Xu_2010} models. We assume that systems with a Hertzsprung gap (HG) donor merge during the CE phase \citep{Belczynski_2007}. This is due to the fact that in {\tt StarTrack} the HG phase begins at the end of the MS and it represents a period of huge expansion for the star. At the beginning of the RLOF phase these donors are often only partially expanded post-MS stars and it is not therefore clear if they already have a well-separated core-envelope structure.  We adopt the weak pulsation pair-instability supernovae and pair-instability supernovae formulation from \cite{Woosley_2017} \& \cite{Belczynski_2016}, which limits the mass spectrum of single BHs to 50\,M$_\odot$. We also use a delayed supernova engine \citep{Fryer_2012,Belczynski_2012}, which affects the birth mass of NSs and BHs so that it allows for the formation of compact objects within the first mass gap ($\sim$ 2 - 5\,M$_\odot$), with the assumption that anything beyond 2.5\,M$_\odot$ is a BH and therefore everything below is a NS \citep{horvath2020}. The metallicity dependence of many \cite{Hurley_2000} formulae is expressed as $\zeta$~=~log(Z
/\,0.02), which represents a normalisation of metallicity $Z$ over solar metallicity $Z_\odot$, which was set at 0.02 . In our simulations we keep solar metallicity at $Z_\odot$~=~0.02 \citep{Grevesse&Sauval_1998} and we set a Maxwellian distribution of natal kicks of $\sigma=265$ km $s^{-1}$\citep{Hobbs_2005}, lowered by fallback during the core-collapse \citep{Fryer_2012} by multiplying it with a fallback factor $f_b$\,$\in$~(0; 1). In order to compare the maximum stellar expansion from our default {\tt StarTrack} prescription with the simulations from the other presented models, we simulate a population of stars with $M_{\rm ZAMS}$ between 18 and 150 M$_\odot$ at $Z$~=~0.1$\,Z_\odot$~=~0.002 . The results of these simulations are shown Figure \ref{fig:RM_Z}.



\subsection{Model 2 -- METISSE-BoOST}
\label{subsec:METISSE-BOOST}

We use the {\tt METISSE} interpolated models from the BoOST \citep{szecsi2021} project. Following \citet{Agrawal_2020}, we use the 'dwarfA' set with metallicity, $Z= 0.00105$. 
The models have been computed using the Bonn Code \citep{Heger_2000, Brott_2011} from ZAMS till the end of CHeB. 
However, the models of massive stars develop numerical instabilities that inhibit computing their evolution after a certain point in 1D stellar evolution codes \citep{Agrawal:2022}. For such models, the remainder of their evolution has been approximated in post-processing (with the so-called "direct extension method" \citealt{szecsi2021}). The models also include a small amount of rotation (at $100$\,km s$^{-1}$). Further input parameters are described in \citet{szecsi2021}. 
The models were fed into the {\tt METISSE} interpolator to create a set of 100 interpolated tracks uniformly distributed in mass between 9 and 100\,M$_\odot$.
Although the initial metallicity differing from what has been used in Model 3 and Model 4, it still falls in the range at which {\tt SSE/StarTrack} predict the highest radii (for more information look at Sec. \ref{subsec:Metallicity}).

\subsection{Model 3 -- METISSE-MESA}
\label{subsec:METISSE-MESA}

These models were computed using the version 11701 of MESA from pre-MS to the end of core carbon burning ($X_c \leq 10^{-4}$) , for metallicity $Z = 0.00142$. In computing these models, mass-loss rates from \citet{Glebbeek:2009} and mixing parameters from \citet{Choi:2016MIST} have been used. The models do not include rotation. To account for eruptive mass loss at super-Eddington luminosity an additional mass loss from super-Eddington wind scheme of MESA reduced by a factor of 10 has been employed whenever stellar luminosity exceeded 1.1 times the mass-weighted average of Eddington luminosity. The convection is modelled following the mixing length theory \citep{Bohm-Vitense_1958}, which is a 1D approximation of the radial motion of blobs of gas over a radial distance that corresponds to the \textit{mixing length} \textit{l$_m$}, after which they dissolve in their surroundings and either release or absorb heat depending on whether the motion was upwards or downwards. In this case the mixing length parameter implemented in {\tt MESA} has been kept at $l_m$ = 1.82 . To suppress the numerical instabilities in massive stars, MLT++ \citep{Paxton2013} scheme of MESA has also been used, which is a treatment for convection that reduces the temperature gradient and therefore the superadiabaticity in some radiative envelope layers of massive stars, so that the simulation time-steps in {\tt MESA} do not get extremely small. This in turn means that the stellar effective temperature increases \citep{Klencki_2020} without altering the actual stellar luminosity, which leads to an inhibition of the stellar radial expansion. Other physical inputs are the same as described in \citet{Agrawal:2021b}.
Following \citet{Agrawal_2020}, the MESA models were interpolated with {\tt METISSE} to create a distribution of stellar tracks, similar to Model 2.

\subsection{Models 4a and 4b -- MESA}
\label{subsec:our_MESA}
For these simulations we used the version 15140 of {\tt MESA} to develop a set of stellar tracks from 18 to 100\,M$_\odot$ at ZAMS that could give the lowest possible maximum stellar radii while still having realistic input physics. For Model 4a these simulations were initialised with the MLT++ physics to artificially reduce the stellar radii during the giant phase, while for Model 4b the MLT++ module was turned off. This is the only difference between the two models.
The input physics has been chosen to enhance the mixing of chemical components inside the various shells of a star (core included), which in turn affects the nuclear reactions and the radial evolution of the object. We adopted the Ledoux criterion for the determination of the convective boundaries \citep{Ledoux_1947}. The metallicity has been set at $Z$= 0.00142 to be consistent with Model 3.\\
\\
The core overshooting leads a star during its MS lifetime to mix the layers above the convective core, therefore increasing the stellar MS lifetime due to a replenishment of the hydrogen reservoirs in the core and in turn rising the nuclear timescale. 
In our models we adopt the step-overshooting approach, whose representing parameter representing is $\alpha_{ov}$, which represents the fraction of the pressure scale height $X_{\rm p}$ by which convective eddies keep travelling up to a distance $l_{ov}$ beyond the convective core boundary \citep{HigginsVink2019}. We set a value of $\alpha_{ov}=0.5$ for our simulations. Our choice for this unusually high value for step-overshooting is motivated by the calibrations of \cite{HigginsVink2019} on HD 166734 and by the 3D hydrodynamics
simulations from \cite{Scott_2021}. In \cite{Gilkis_2021} values up to 1.2 were also explored for ZAMS masses up to $\sim$ 200\,M$_\odot$.\\
\\
Semiconvection is defined as the mixing in regions that are unstable to the Schwarzschild criterion for convention, but stable to the Ledoux one. This means semiconvective zones are defined as regions where the molecular composition gradient is higher than the temperature gradient, which is in turn higher than the adiabatic one. As described in \cite{Paxton2013}, these regions experience mixing through a time-dependent diffusive process. After the end of core hydrogen burning, the following contraction phase of massive stars leads them to ignite the hydrogen in a shell. Due to this reason the deep hydrogen envelope forms semiconvective regions inside of it. In this context the diffusion coefficient is directly proportional to the factor $\alpha_{sc}$, which is a dimensionless parameter that describes the mixing efficiency. \cite{Schootemeijer_2019} has shown that for non-rotating models the value of $\alpha_{sc}$ could be realistically up to 300, which we chose for our simulations. This impact on internal mixing is however reduced in case of a very efficient core overshooting, since if a considerable part of the stellar hydrogen reservoirs is dragged inside the stellar core during the MS due to efficient mixing, it will be harder for semiconvective regions of localised hydrogen burning to form during the core He burning phase \citep{Schootemeijer_2019}. With very low or null values of overshooting semiconvective regions can already form during the MS phase. As we expected, from our checks we found that semiconvection did not have an important role in the simulations, considering the adopted $\alpha_{ov}$.\\
\\
The wind prescription that was adopted in our {\tt MESA} simulations is the so-called Dutch wind prescription \citep{Glebbeek:2009}, which uses different wind models depending on the effective temperature $T_{\rm eff}$ and the surface hydrogen abundance $H_{\rm sur}$ (in terms of the fraction of hydrogen on the stellar surface). As shown in Table~\ref{tab.dutch}, if $T_{\rm eff}<10^4$~K the \citet{deJager1988} model is used regardless of the hydrogen surface abundance. When $T_{\rm eff}>10^4$ K, instead, the Dutch prescription either adopts the \citet{NugisLamers2000} model ($H_{\rm sur}<0.4$), or the \cite{Vink2001} model ($H_{\rm sur}>0.4$).

\begin{table}
\caption{Dutch Stellar Winds prescription in {\tt MESA}.}
\centering
\begin{tabular}{c c c}
\hline\hline
   & \textbf{$T_{\rm eff}<10^4$ K} & \textbf{$T_{\rm eff}\geq10^4$ K} \\ 
\hline
 --                     & \citet{deJager1988} & -- \\
 $X_{\rm sur}<0.4$      & -- & \cite{NugisLamers2000}  \\ 
 $X_{\rm sur} \geq 0.4$ & -- & \citet{Vink2001} \\ 
\hline
\end{tabular}
\label{tab.dutch}
\end{table}

The \cite{deJager1988} winds are described by the following equation:
\begin{equation}
 \log(\dot{M}) = 1.769 \log\left(\frac{L}{{\rm L}_\odot}\right) - 1.676 \log(T_{\rm eff}) - 8.158
\end{equation}

where $\dot{M}$ is in ${\rm M}_\odot yr^{-1}$. The wind mass loss for Wolf-Rayet stars follows instead the \citet{NugisLamers2000} model:  
\begin{equation}
  \dot{M} = 10^{-11}\left(\frac{L}{{\rm L}_\odot}\right)^{1.29}Y^{1.7}\sqrt{Z}
\end{equation}

where Y is the He surface abundance and Z the metallicity.

\section{Results}
\label{sec:ResDisc}

\subsection{Dependence of the maximum radius on mass and metallicity}
\label{subsec:Metallicity}
\begin{figure}
    \centering
	    \includegraphics[width=\columnwidth]{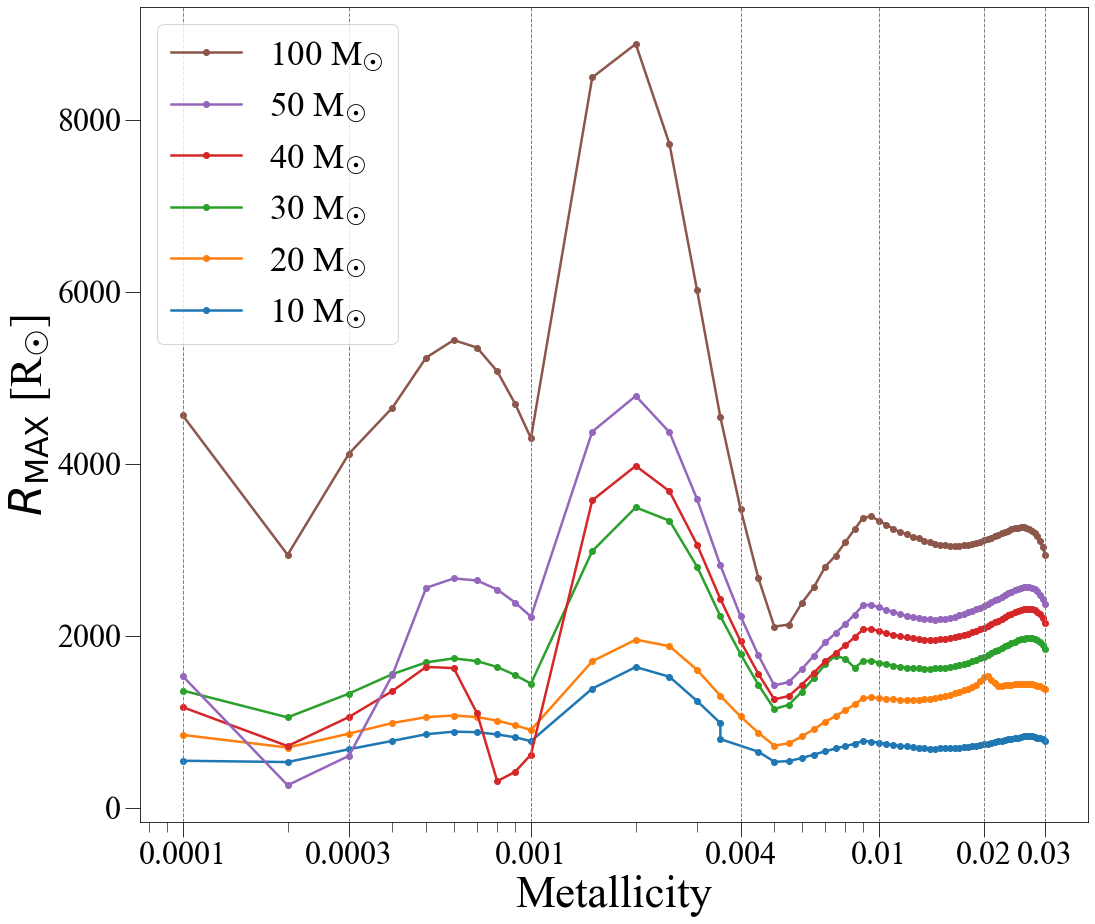}
        \caption{Maximum radii of massive stars as a function of their initial metallicity from standard {\tt StarTrack} simulations. Each line corresponds to a different ZAMS mass between 10 and 100\,M$_\odot$. The vertical lines correspond to the original stellar set of models from \citealt{Pols_1998} from which the analytic \citealt{Hurley_2000} formulae were fitted. Artefacts due to extrapolation and interpolation of the original evolutionary equations could be the cause for the absence of a quasi-linear relation between $R_{\rm MAX}$ and metallicity.}
        \label{fig:RM_Z}
\end{figure}

In Figure \ref{fig:RM_Z} we show the dependence of the maximum stellar radius on $M_{\text{ZAMS}}$ and metallicity, as simulated with {\tt StarTrack}. Each line represents a different ZAMS mass: 10, 20, 30, 40, 50 and 100\,M$_\odot$. The vertical grey lines show the initial metallicities of the simulations from \cite{Pols_1998}, upon which the models from \cite{Hurley_2000} were built: 0.0001, 0.0003, 0.001, 0.004, 0.01, 0.02, 0.03 .

\begin{figure}
    \centering
	    \includegraphics[width=\columnwidth]{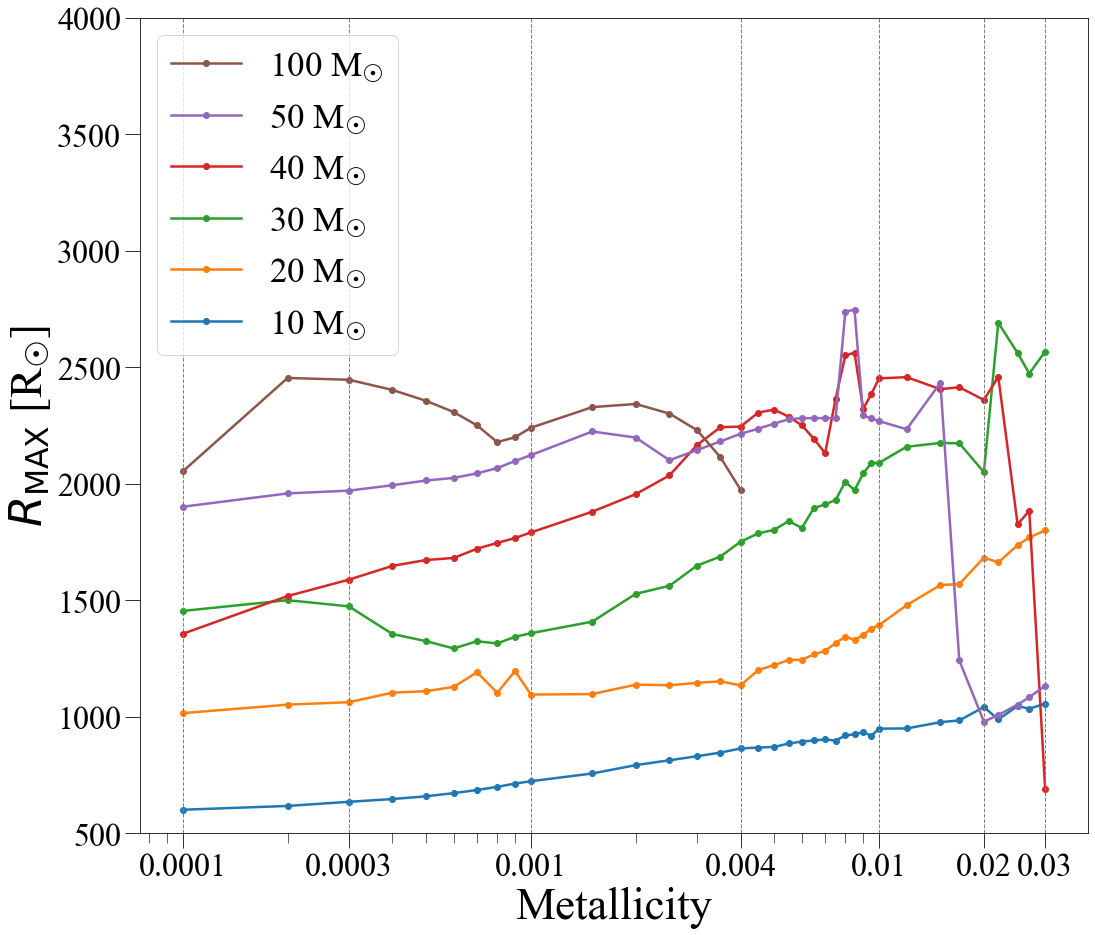}
        \caption{Maximum radii of massive stars as a function of their initial metallicity from our Model 4b {\tt MESA} simulations. Each line corresponds to a different ZAMS mass between 10 and 100\,M$_\odot$. $R_{\rm MAX}$ almost linearly increases as a function of metallicity due to increased opacity levels.}
        \label{fig:RM_Z_MESA}
\end{figure}

In general the stellar radius is dependent on the ZAMS mass and on the initial metallicity of the object. On the other hand stellar winds are proportional to metallicity and therefore may significantly reduce the mass and the radius of a star. In \cite{Hurley_2000} formulae, as implemented in the {\tt StarTrack} code, the maximum radial expansion always peaks at around $Z$= 0.002 . It must be pointed out that the results around this metallicity are just interpolations from the \cite{Hurley_2000} fits of the stellar set of models from \cite{Pols_1998}. The maximum radius peak could therefore be only an artefact. Also, the evolution of any star with a ZAMS mass beyond 50\,M$_\odot$ is an extrapolation of the formulae, since they were only fit for stars up to that initial mass. Likewise, the stellar set of models from \cite{Pols_1998} was computed for ZAMS masses between 0.5 and 50\,M$_\odot$ with a spacing of $\sim$ 0.1 in log$M$, which means that for any other initial mass within that range the stellar evolution is also interpolated. Such an interpretation could be strengthened by our Model 4b {\tt MESA} simulations (see Sec. \ref{subsec:our_MESA}) shown in Figure \ref{fig:RM_Z_MESA}, which we repeat is the model that doesn't adopt the MLT++ scheme to treat superadiabatic radiative envelopes layers of massive stars. The line for 100\,M$_\odot$ stops at a metallicity of 0.004 because beyond that point with our {\tt MESA} setup the simulations reach a timestep limit that makes it impossible to get past the main sequence. In these {\tt MESA} calculations the maximum stellar radius almost always increases with metallicity for stars below 40\,M$_\odot$. For the most massive stellar tracks the maximum radius increases less steeply as a function of metallicity if compared to lower mass stars. This is a result of an increased internal mixing for massive stars that reduces the mass of the H-rich envelope limiting the radial expansion beyond main sequence. Additionally, for larger metallicities stellar winds become strong enough to help effectively remove the H-rich envelope and further reduce stellar expansion (see the large decrease in radius for the $M_{\text{ZAMS}}$~=~40~M$_\odot$ and $M_{\text{ZAMS}}$~=~50~M$_\odot$ models around $Z$~=~0.02). 
For 30\,M$_\odot$ at $Z$\,=\,0.02 and 40\,M$_\odot$ at $Z$\,=\,0.025, we take $R_{\rm MAX}$ before the radial behaviour of models is dominated by numerical noise due to proximity to the Eddington limit. This was shown in Sec. 7.2 of \cite{Paxton2013} to be a well-known issue for detailed simulations of massive and high-metallicity post-MS stars. Since the numerical difficulties occur in the red supergiant stage and typically only after the star has began moving leftwards in the HR diagram (i.e. contracting), we deem that taking their radii just before that time as their maximum expansion is an optimal approximation, considering also that the loosely bound envelopes lead to drastic mass loss events that keep the stars from further expanding. It is important to highlight that, on the contrary to our {\tt StarTrack} simulations, our detailed simulations do not show $R_{\rm MAX}$ peaking at $Z$= 0.002 (or at any other initial metallicity) for every $M_{\rm ZAMS}$.

\subsection{Models fit}
\label{subsec:models_fit}

For all the models we fitted a logarithmic behaviour for $R_{\text{MAX}}$ [R$_\odot$] as a function of $M_{\text{ZAMS}}$ [M$_\odot$]. The following formulae are labelled as log$R_{\text{MAX,1}}$, log$R_{\text{MAX,2}}$, log$R_{\text{MAX,3}}$, log$R_{\text{MAX,4a}}$ and log$R_{\text{MAX,4b}}$ to indicate which model they were fit from. For $M_{\text{ZAMS}}$<18\,M$_\odot$ the maximum radius has not been constrained by the following formulae and the stars evolve according to our standard evolutionary prescription.
For \textbf{Model 1} we retrieved the following logarithmic equation:

\begin{equation}
     {\rm log}R_{\rm MAX,1} = 0.1\times{\rm log}(3177.1M_{\text{ZAMS}}) + 0.006M_{\text{ZAMS}} + 2.8
\end{equation}

It must be stressed that we do not use Model 1 to constrain the maximum stellar radius, since this already represents our default prescription. This formula was added as a further comparison with the other models.

For \textbf{Model 2}, we fitted the following logarithmic relation for $R_{\text{MAX}}$:

\begin{equation}
  \begin{aligned}[b]
     &{\rm log}R_{\text{MAX,2}} = 1.358\times {\rm log}(2.712\times10^{-13}M_{\text{ZAMS}})\\
     &-6.536\times10^{-3}M_{\text{ZAMS}}+18.443
    \end{aligned}
\label{eq:Mod2}
\end{equation}

For ZAMS masses higher than 100\,M$_\odot$ the maximum radius is set at $R_{\text{MAX,2}}$(100 ${\rm M}_\odot)\sim 2752$\,R$_\odot$.\\
\\
Considering the behaviour of the {\tt METISSE-MESA} simulations in \textbf{Model 3}, we divided the simulation results into $M_{\text{ZAMS}}$ bins and fitted each segment separately. The results of our fits can be found in Table~\ref{tab:MESAfit}. Every star with a ZAMS mass beyond $100$\,M$_\odot$ is by default set at $R_{\rm MAX,3}$(100 ${\rm M}_\odot)\sim631$\,R$_\odot$.

\begin{table}
	\centering
	\caption{Fitted {\tt METISSE-MESA} formulae}
	\label{tab:METISSE_MESA}
	\begin{tabular}{lcr} 
		\hline
		$M_{\text{ZAMS}}$ [M$_\odot$] & log$R_{\text{MAX,3}}$ =\\
		\hline\hline
		18-20 & log$R_{\text{MAX,2}}$\\
        20-40 & log$R_{\text{MAX,2}}$(20 {\rm M}$_\odot$)+0.0002$M_{\text{ZAMS}}$\\
        40-45 & (-9$M_{\text{ZAMS}}$+510)/50\\
        45-52 & (-$M_{\text{ZAMS}}$+321)/20\\
        52-71 & (3$M_{\text{ZAMS}}$+47)/100\\
        71-100 & (2$M_{\text{ZAMS}}$+612)/290\\
        $\geq$100 & 2.8\\
		\hline
	\end{tabular}
	\label{tab:MESAfit}
\end{table}

$R_{\text{MAX}}$ for our {\tt MESA} simulations in \textbf{Model 4a} is described as:

\begin{equation}
\begin{aligned}[b]
     &{\rm log}R_{\text{MAX,4a}} = -11.591\times{\rm log}(2.4341M_{\text{ZAMS}}) + 0.329M_{\text{ZAMS}}\\
     &-3.183M_{\text{ZAMS}}^2\times10^{-3}+1.197M_{\text{ZAMS}}^3\times10^{-5} + 17.037
\end{aligned}
\label{eq:Mod4a}
\end{equation}

Since the relation is only valid for ZAMS masses between 18 and 100\,M$_\odot$, therefore for stars beyond the upper limit we set $R_{\text{MAX,4a}}$($M_{\text{ZAMS}} > 100$ ${\rm M}_\odot)$ = $R_{\text{MAX,4a}}$(100 ${\rm M}_\odot)\sim322$\,R$_\odot$.\\
\\
Finally, for our \textbf{Model 4b} we fitted this behaviour for ZAMS masses between 18 and 100\,M$_\odot$:

\begin{equation}
\begin{aligned}[b]
     &{\rm log}R_{\text{MAX,4b}} = 1.143\times{\rm log}(2.767\times10^2M_{\text{ZAMS}})\\
     &-5.825\times10^{-3}M_{\text{ZAMS}}-1.129
\end{aligned}
\label{eq:Mod4b}
\end{equation}

\begin{figure*}
	\includegraphics[width=18 cm]{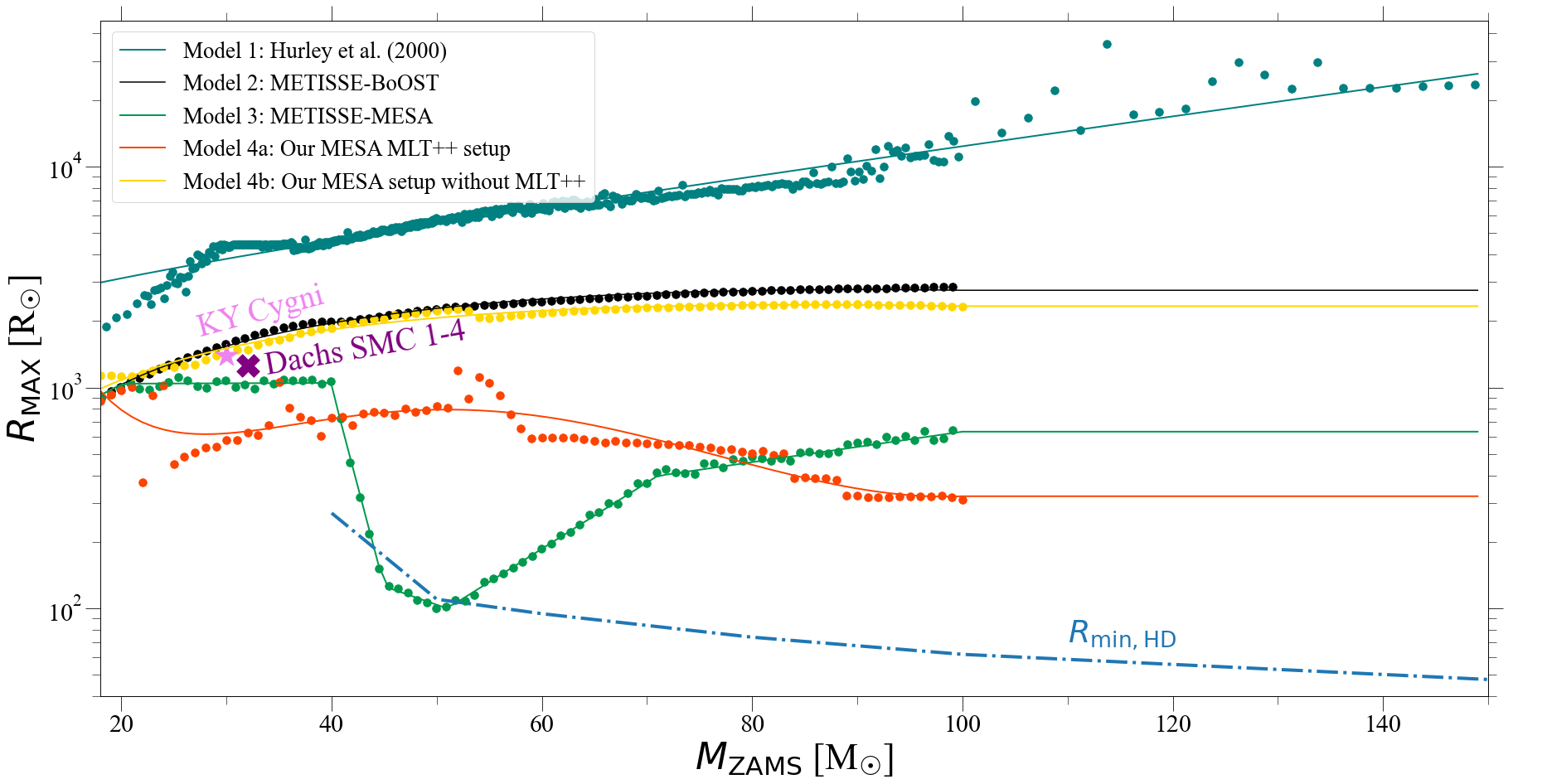}
    \caption{Maximum stellar radii as a function of $M_{\rm ZAMS}$ for each presented model. We show the maximum stellar radii obtained with the \citealt{Hurley_2000} rapid evolutionary formulae used in many codes (e.g. {\tt StarTrack}, {\tt COMPAS}, {\tt MOCCA}), and the ones obtained from Models 2, 3, 4a and 4b. The dots are the data points from the estimates from {\tt StarTrack} (Model 1)  or detailed calculations (Model 2, 3, 4a, 4b). They are the same colour of the lines representing the $R_{\rm MAX}$ prescriptions. As a reference we show the radius of KY Cygni ($\sim$ 1500\,R$_\odot$, for $M_{\rm ZAMS}$ = 30\,M$_\odot$) and Small Magellanic Cloud Dachs 1-4 ($\sim$ 1300\,R$_\odot$  $M_{\rm ZAMS}$ = 32\,M$_\odot$; see Sec. \ref{subsec:models_fit} for details). We also show with the dotted line the minimum radial expansion as a function of $M_{\rm ZAMS}$ that a star must reach to cross the Humphrey-Davidson limit according to our simulations. With the partial exception of Model 3, all of the proposed $R_{\rm MAX}$ models cross this limit by even orders of magnitude. Maximum stellar radii for the same $M_{\rm ZAMS}$ may differ by more than one order of magnitude depending on which code and input physics is used.
    }
    \label{fig:RMAX}
\end{figure*}

For more massive stars $R_{\text{MAX,4b}}$($M_{\text{ZAMS}} > 100$ ${\rm M}_\odot)$ = $R_{\text{MAX,4b}}$(100 ${\rm M}_\odot)\sim2332$\,R$_\odot$. Despite the two models were taken from different evolutionary codes, Model 2 and Model 4b predict similar logarithmic behaviours of the maximum expansion as a function of $M_{\rm ZAMS}$. For $M_{\text{ZAMS}} > 100$ ${\rm M}_\odot$, $R_{\text{MAX,2}}$ and $R_{\text{MAX,4b}}$ differ of a factor $\sim$1.2 .\\ 
\\
In Figure \ref{fig:RMAX} we plotted the maximum radius that stars can reach during their whole lifetime as a function of their ZAMS masses. The solid lines represent log$R_{\text{MAX,1}}$, log$R_{\text{MAX,2}}$, log$R_{\text{MAX,3}}$, log$R_{\text{MAX,4a}}$ and log$R_{\text{MAX,4b}}$ as a function of $M_{\rm ZAMS}$, while the dots are the data points that have been used to fit the logarithmic equations (they are the same colour of their respective model line). 
If the radial evolution for stars with a $M_{\text{ZAMS}}\geq$ 18\,M$_\odot$ is simulated with our standard Model 1 prescription, it can reach a maximum value between 2$\times 10^3$ and 3.6$\times 10^4$\,R$_\odot$. The differences in terms of $R_{\rm MAX}$ depend on $M_{\rm ZAMS}$ and they increase with increasing initial masses. For low initial masses ($M_{\rm ZAMS}\lesssim$ 40\,M$_\odot$) the detailed models give maximum radii smaller by a factor of $\sim$2 (Model 2), $\sim$3 (Model 3), $\sim$5 (Model 4a) and $\sim$3 (Model 4b) than the ones from \cite{Hurley_2000} formulae (Model 1). For high initial masses ($M_{\rm ZAMS}$ $\gtrsim$ 100\,M$_\odot$)  the detailed models give maximum radii smaller by a factor of $\sim$7 (Model 2), $\sim$30 (Model 3), $\sim$60 (Model 4a), $\sim$8 (Model 4b). It must be highlighted that Model 2 and Model 4b, which were built from different evolutionary codes ({\tt BoOST} and {\tt MESA}), without any prior calibration to make them produce similar results, predict similar trends of maximum radial evolution as a function of ZAMS mass. We also show in this plot $R_{\rm min, HD}$ , which is the minimum radial expansion as a function of $M_{\rm ZAMS}$ at which our simulated stars cross the Humphrey-Davidson limit (defined in Eq. \ref{eq:HDlimit}) according to Model 1. The lowest $M_{\rm ZAMS}$ in our sets at which a stellar track crosses the HD limit is $\sim$40\,M$_\odot$, hence the starting point for the $R_{\rm min, HD}$ line on the left hand side of Fig. \ref{fig:RMAX}.

\begin{figure}
	\includegraphics[width=\columnwidth]{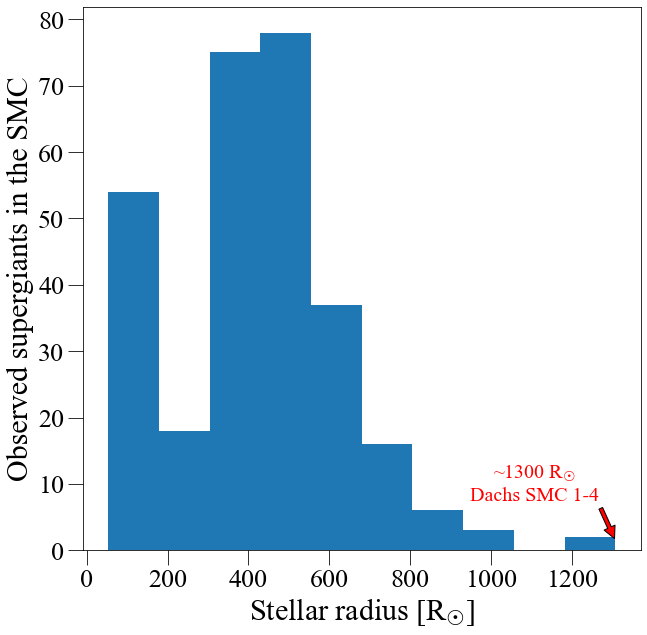}
    \caption{Estimated radii of RSGs, YSGs, and BSGs in the Small Magellanic Cloud. The largest one is Dachs SMC 1-4 at $\sim$1300\,R$_\odot$, as shown by the red arrow.}
    \label{fig:Obs}
\end{figure}

As a sanity check we also compared the $R_{\text{MAX}}$ in our models with all the Red, Yellow and Blue Supergiants (respectively RSGs, YSGs, and BSGs) in the Small Magellanic Cloud (SMC, Figure \ref{fig:Obs}), with a particular focus on the coolest and most luminous supergiants.  We use the same compilation of objects as \citet{Gilkis_2021}. The sample includes all known cool supergiants with estimated luminosities of $\log L > 4.7\,[\ergs]$ and effective temperatures smaller than $T_{\rm eff} < 12\,500$\,K, which probes the horizontal edge of the Humphreys-Davidson limit \citep{Humphreys1979}.  The catalogue compiles RSGs and YSGs from \citet{Davies2018}, \citet{Neugent2010}, and BSGs that are found from Gaia DR2 \citep{GaiaDR2} using colour-$T_{\rm eff}$ calibrations from \citet{Evans2003} or derived properties from \citet{Dufton2000}. The sample was cleaned from non-SMC contaminants using proper motion and parallax criteria.  The total list comprises 179 stars: 140 RSGs, 7 YSGs, and 32 cool BSGs. We refer to \citet{Gilkis_2021} for more details.\\
\\
To retrieve the radial extension of the supergiants in our sample, we used the Stefan–Boltzmann law:

\begin{equation}
    \frac{L}{L_\odot} = \left(\frac{T}{T_\odot}\right)^4 \left(\frac{R}{R_\odot}\right)^2
\end{equation}

\begin{figure}
	\includegraphics[width=\columnwidth]{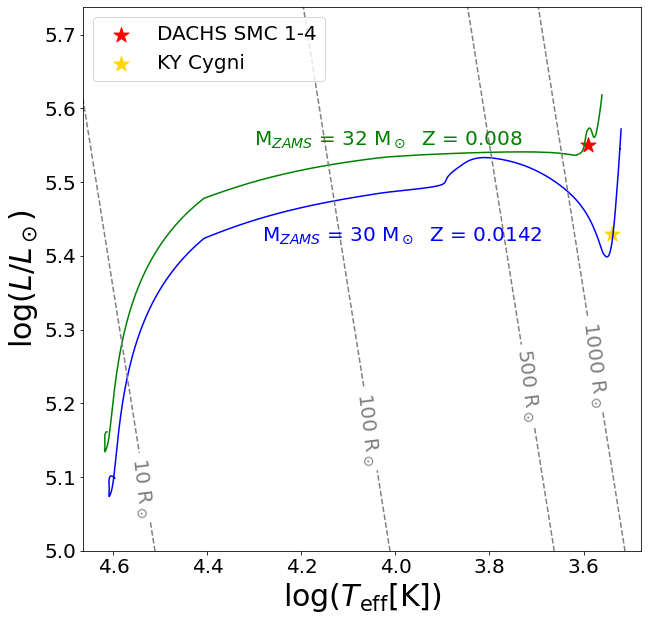}
    \caption{H-R diagram of two stars as simulated by our {\tt MESA} setup from Model 4b. As a reference we show the position of Dachs SMC 1-4 and KY Cygni in terms of luminosity and effective temperature.}
    \label{fig:HRAll}
\end{figure}

The largest star known in the SMC is Dachs SMC 1-4, a RSG with a radius of $\sim1300$\,R$_\odot$, a luminosity of log($L$/L$_\odot$) = 5.55 and log($T_{\rm eff}$ [K]) = 3.59. 
With the {\tt MESA} setup from Model 4b, an initial metallicity of at $Z$ = 0.008, we tested different $M_{\rm ZAMS}$ values to check with which evolutionary track we could get the same position of Dachs SMC 1-4 in the H-R diagram. We found an ideal candidate in a star of $M_{\rm ZAMS}$~=~32~M$_\odot$, with log($T_{\rm eff}$) $\sim$ 3.60, log($L/{\rm L}_\odot$) $\sim$ 5.55 and $R$~$\sim$~1260~R$_\odot$. The grey dashed lines represent where, for a specific luminosity and effective temperature, a star has a radius of 10, 100, 500 and 1000\,R$_\odot$. Also KY Cygni, one of the largest stars in the Milky Way, has been used for this comparison, since it can even go beyond the expansion of Dachs SMC 1-4 and it has been estimated to be $\sim$ 1500\,R$_\odot$ large \citep{Levesque2005}, with log($L/{\rm L}_\odot$) = 5.43 and log($T_{\rm eff}$) = 3.54 \citep{Dorn_Wallenstein_2020}. Like in the case with Dachs SMC 1-4, we used the same setup to find an evolutionary track that could match KY Cygni in the H-R diagram, but at a metallicity of 0.0142 (solar metallicity). With log($T_{\rm eff}$) = 3.53, log($L/{\rm L}_\odot$) = 5.44 and $R$ = 1500.44\,R$_\odot$ we found an optimal candidate in a star of $M_{\rm ZAMS}$ = 30\,M$_\odot$. The HR position of both stars and their respective simulated tracks are shown in Fig. \ref{fig:HRAll}. The KY Cygni and the Dachs SMC 1-4 radii as a function of their estimated ZAMS masses are shown in Fig. \ref{fig:RMAX}.

As a reference for the $R_{\rm MAX}$ prescriptions, we show the estimated radii of Dachs SMC 1-4 and KY Cygni from the literature, in order to test if our models could underestimate the real radial expansion of stars. This comparison shows that Model 1, Model 2 and Model 4b can reproduce radii that are high enough to be compatible with the given observational constraints, while Model 3 and Model 4a can reach at most 1050\,R$_\odot$. It must be pointed out that our $R_{\text{MAX}}$ prescriptions are only as a function of $M_{\rm ZAMS}$: they do not take into account any other factor like metallicity or the assumed mixing length. Estimations of RSGs radii must also be taken with a grain of salt, considering the margins of error that come along their luminosities and effective temperatures. 

\subsection{Estimated LVK local merger rate densities}
\label{sec:LVKrates}


In Table~\ref{tab:MRD} we compare the merger rate densities as reported by the LVK collaboration for a fiducial redshift of $z\sim0.2$ with the ones we calculated for the same redshift range.\\
\\
Model 2, whose $R_{\rm MAX}$ as a function of $M_{\rm ZAMS}$ is between 2 and 7 times smaller than what is obtained from Model 1, differs only by 3 per cent from the standard prescription in terms of BH-BH merger rate density. We explain that this is due to the fact that the rate of RLOF events for BH-BH merger progenitors remains unaltered from Model 1, since the simulated stars in BH-BH progenitors overflow their Roche lobe before reaching their respective $R_{\text{MAX}}$ (for more details see Sec. \ref{sec:RLOF_env}).
\\
\\
Model 3, which predicts maximum radii between 3 to 30 times smaller than Model 1, does show a decrease of the BH-BH merger rate density of $\sim$30 per cent, with an increase of $\sim$25 per cent in the BH-NS merger rate density and the same NS-NS merger rate density (within the numerical variability of population synthesis estimations) of Model 1. This is due to the fact that with this model more stars in BH-BH merger progenitors reach their $R_{\rm MAX}$ before overfilling their Roche lobe. For BH-NS progenitors, instead, many RLOF events that bring a close binary to an early merger event are not initialised in the first place, which brings to a higher survival rate for binaries and in turn to an increase in the number of BH-NS mergers.\\
\\
Model 4a, which gives the smallest values of $R_{\rm MAX}$ for ZAMS masses below $\sim40$\,M$_\odot$ and above $\sim85$\,M$_\odot$, shows a decrease from Model 1 in the merger rate densities for all the channels. The BH-BH merger rate density is 18.7 Gpc$^{-3}$yr$^{-1}$, which is the lowest one among our models. In this scenario the BH-BH merger rate density show a decrease by a factor $\sim$3.5. The NS-NS and BH-NS merger rate densities are respectively roughly 40 and 35 per cent lower than the ones from Model 1.
This shows that with this model the reduced rate of initialisation of RLOF events is reducing the number of close double compact objects that lead to mergers within a Hubble time, rather than forming more close double compact objects due to a reduced number of early mergers.\\
\\
Finally, with the Model 4b prescription, we predict merger rate densities for BH-BH and NS-NS binaries that are fully compatible with both Model 1 and Model 2. The only exception being the BH-NS merger rate density, which is the highest one among the models.\\
\\
With these results we show that our standard prescription (Model 1) is as reliable in terms of double compact object merger rate density estimates from isolated binary evolution as the other models considered here, which predict a much more restrictive radial evolution. This is motivated by the fact that $R_{\text{MAX}}$ in our prescription was reduced sometimes more than one order of magnitude from Model 1 and that the lowest BH-BH merger rate density estimation (Model 4a) was roughly three times lower than the one from Model 1. This conclusion is also strengthened by the fact that, as previously mentioned in Sec. \ref{sec:intro}, merger rate density estimations have a way wider margin of variability than what has been shown in this study.\\
\\
Other two important results to highlight are (i) most of the RLOF events that lead to BH-BH mergers happen below the radial maximum described by Model 4b (ii) for redshifts at 0.2 all our models besides Model 4a estimate a BH-BH merger rate density that is beyond the reported LVK 90 per cent credibility range. On the other hand, considering the uncertainties in population synthesis studies, slight variations in the input physics for Model 1, Model 2, Model 3 and Model 4b in the isolated binary channel could make their BH-BH merger rate estimations to fit in the LVK credibility range (see e.g. \citealt{Dominik_2012}).

\begin{table}
	\centering
	\caption{Comparison between the LIGO/Virgo/KAGRA local ($z$ $\sim$0.2) merger rate densities \citep{LIGO_2021,Abbott2021} [Gpc$^{-3} $yr$^{-1}$] and the ones calculated with our models within a redshift of 0.2. With the bold font we mark the model for which BH-BH, BH-NS and NS-NS merger rate densities are within the LVK 90 per cent margin of credibility.}
	\begin{tabular}{lccr} 
		\hline  
		Model & BH-BH & BH-NS & NS-NS\\
		\hline\hline
        LVK & 17.9 - 44 & 7.8 - 140 & 10 - 1700\\ 
		\hline
		Model 1 & 68.1 & 15.6 & 158.9\\ 
		Model 2 & 65.8 & 14.9 & 162.8\\
		Model 3 & 46.6 & 19.7 & 157.7\\
		\textbf{Model 4a} & \textbf{18.7} & \textbf{9.5} & \textbf{100.4}\\
		Model 4b & 65.6 & 22.9 & 160.8\\
		\hline
	\end{tabular}
	\label{tab:MRD}
\end{table}

\subsection{Estimated BH-BH mass distribution}
\label{subsec:mass_distr}

\begin{figure*}
	\includegraphics[width=18 cm]{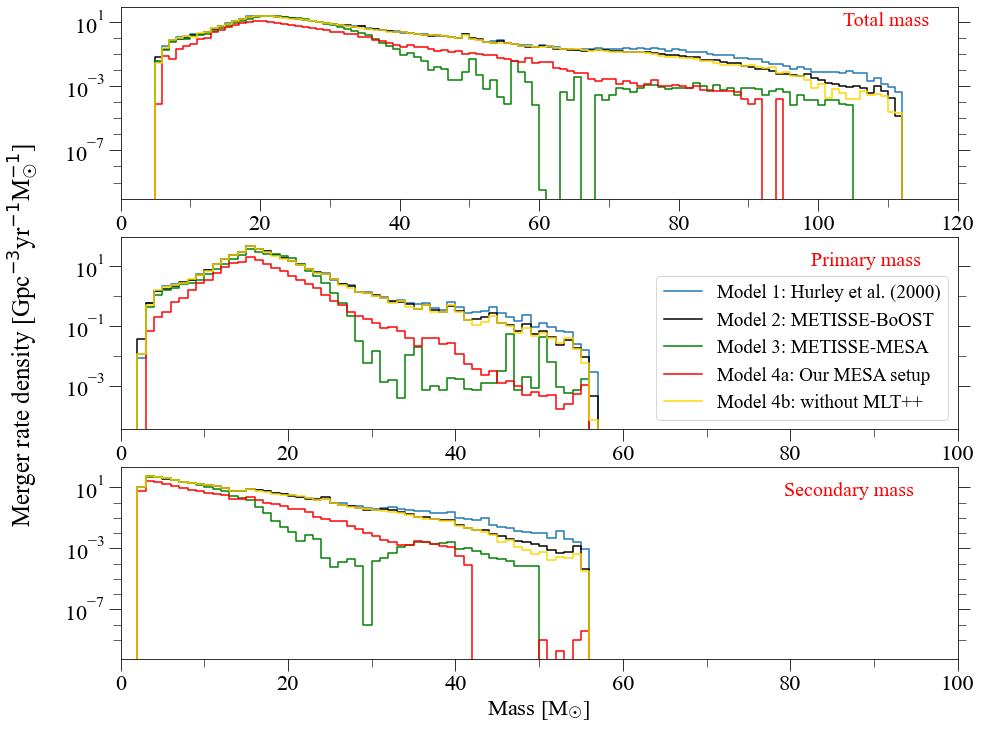}
    \caption{BH-BH mass distribution for mergers within redshift $z$= 2. The plot on the top represents the merger rate density as a function of the total BH-BH mass. The one in the middle the mass distribution for the primary (i.e. the most massive) BH, while the plot on the bottom shows the mass distribution of the secondary BH. Similarly to Fig. \ref{fig:RMAX} each model is described with a different line.}
    \label{fig:BHBHMass}
\end{figure*}

Fig. \ref{fig:BHBHMass} shows our estimated merger rate density of BH-BH mergers within a redshift $z$= 2 as a function of the total binary mass, the mass of the primary/more massive BH, and the mass of the secondary/less massive BH for all the models described in Sec. \ref{sec:Method} .

Model 2 and Model 4b do not alter the estimates of the mass distribution of BH-BH mergers within redshift z = 2 from our standard setup (Model 1), despite the $R_{\text{MAX}}$ decrease. Model 3 and Model 4a, which are the ones with the smallest $R_{\rm MAX}$, still show mass distributions peaked at 19\,M$_\odot$ (total mass), 15\,M$_\odot$ (primary mass) and 4\,M$_\odot$ (secondary mass). Our estimates show that Model 3 and Model 4a $R_{\rm MAX}$ prescriptions diverge the most (from Model 1 and each other) for a total BH mass higher than 50\,M$_\odot$. This is expected, since, as noticeable in Fig. \ref{fig:RMAX}, the differences between the models in terms of $R_{\rm MAX}$ grow in function of $M_{\rm ZAMS}$. None of our models are therefore in agreement with the reported LVK BH-BH mass distribution in \cite{LIGO_2021}, where two peaks are observed at primary masses of $\sim$10\,M$_\odot$ and $\sim$35\,M$_\odot$.\\
\\
Considering that beyond a total BH mass of 50\,M$_\odot$ the BH-BH merger rate density differs of roughly one order of magnitude between Models 1, 2 and 4b, and Models 3 and 4a, we show in Table \ref{tab:MRD_50Msun} how the same estimates from Table \ref{tab:MRD} can vary for the BH-BH channel if only the mergers with a total mass > 50\,M$_\odot$ are considered.

\begin{table}
	\centering
	\caption{Estimated BH-BH merger rate densities [Gpc$^{-3} $yr$^{-1}$] for total masses beyond 50\,M$_\odot$ for all our models.}
	\begin{tabular}{lccccr} 
		\hline  
		redshift & Model 1 & Model 2 & Model 3 & Model 4a & Model 4b\\
		\hline\hline
		< 0.2 & 2.8 & 2.35 & 0.06 & 0.13 & 2.21\\
		< 2 & 16.3 & 13.2 & 0.3 & 1.1 & 12.8\\
		\hline
	\end{tabular}
	\label{tab:MRD_50Msun}
\end{table}

For this mass range there is a decrease of $\sim$ 16 per cent from Model 1 to Models 2 and 4b for both z$\sim$0.2 and z$\sim$2. Since these BH-BH progenitors are initially very massive stars that are in the $M_{\rm ZAMS}$ range where the predicted $R_{\rm MAX}$ differs the most between the detailed models and Model 1. Model 3 predicts a BH-BH merger rate density that is between $\sim$ 45 (z$\sim$0.2) and $\sim$ 54 (z$\sim$2) times smaller than what is predicted with Model 1, and 2 times smaller than what we estimate with Model 4a.

\subsection{CE events and nature of the stellar envelope for BH-BH merger progenitors}
\label{sec:RLOF_env}
 
\begin{figure*}
	\includegraphics[width=18 cm]{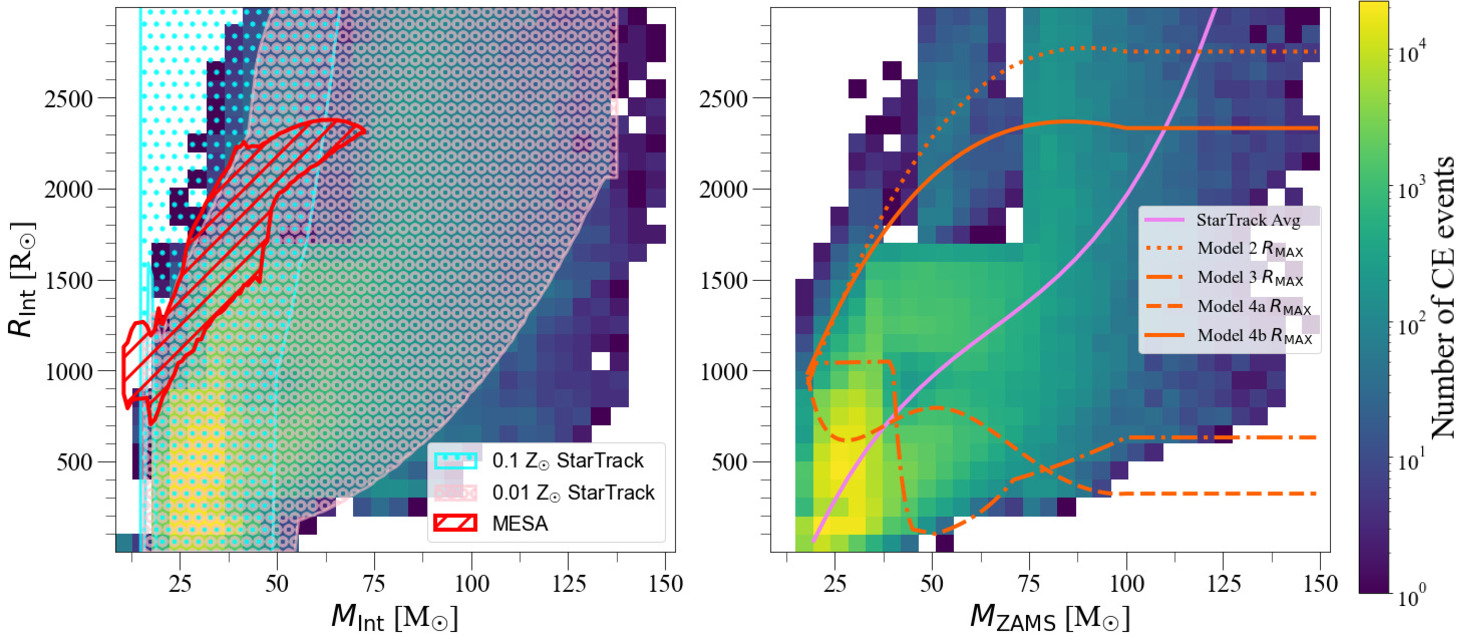}
    \caption{2D histogram showing how often the CE events happen as a function of $R_{\text{Int}}$ with $M_{\text{Int}}$ (left panel) and $M_{\text{ZAMS}}$ (right panel) for binaries that evolve into BH-BH with a merger time withing the age of the Universe. Both plots were rescaled to include only the CE events with $R_{\text{Int}}$ < 3000\,R$_\odot$. On the left plot the red dashed area shows where stars below 100\,M$_\odot$ get a convective envelope according to our {\tt MESA} simulations, while the dotted teal and pink areas show how our {\tt StarTrack} prescription defines CE survival for respectively Z = 0.1 Z$_\odot$ and Z = 0.01 Z$_\odot$. On the right plot the orange lines show the $R_{\text{MAX}}$ as a function of $M_{\text{ZAMS}}$ from our Model 2, Model 3, Model 4a and Model 4b simulations, while the pink line represents the average $R_{\text{Int}}$ values as a function of $M_{\text{ZAMS}}$ (Eq. \ref{eq:Rint_ZAMS}).
    }
    \label{fig:CE_conv}
\end{figure*}

In this section we take an in-depth look at the nature of donor stars in CE events that lead to the formation of BH-BH mergers in our {\tt StarTrack} simulations.\\
\\
According to \cite{Klencki_2020,Klencki_2021} and \cite{Marchant_2021}, CE events initiated by stars without a deep outer convective envelope lead to stellar mergers rather than successful CE ejections. \cite{Gallegos_2021} also showed that population synthesis models can overestimate the number of CE survival rates and underestimate merger times, if compared to detailed binary evolution. From this starting point we used Model 4b to check the nature of the stellar envelope of massive stars. In our methodology we define a deep convective envelope as an envelope that has at least 10 per cent of its mass in the outer convective zone (if any). Note that our {\tt MESA} simulations are limited to the 0.1 Z$_\odot$ metallicity, but the radius above which a star develops an outer convective envelope may depend slightly on metallicity \citep{Klencki_2020}. As a comparison we also did the same simulations with our {\tt StarTrack} setup for metallicities of Z$=0.1$ Z$_\odot$ and Z$=0.01$ Z$_\odot$ for a $M_{\rm ZAMS}$ range up to 150\,M$_\odot$\footnote{Due to a potential extrapolation artefact in {\tt StarTrack}, we report a metallicity-dependent maximum mass beyond which no donor star, according to our default CE survival prescription, survives a CE phase. For Z$=0.1$ Z$_\odot$ this threshold mass is $\sim$ 95\,M$_\odot$, while for Z$=0.01$ Z$_\odot$ this value increases to $\sim$ 135\,M$_\odot$.}. For {\tt StarTrack} the CE survival conditions are described in \cite{Belczynski_2008}. We define the Interaction Radius ($R_{\text{Int}}$) and Interaction Mass ($M_{\text{Int}}$) as the stellar radius and mass at which a donor star in a binary overflows its Roche lobe and initiates a mass transfer event. Both of them are in solar units.
We show the results in Fig. \ref{fig:CE_conv}, where the 2D histograms in the background represent the $R_{\text{Int}}$ and either $M_{\text{Int}}$ (left) or $M_{\text{ZAMS}}$ (right) distribution for CE events that lead massive binaries to form BH-BH systems with a merger time within the age of the Universe, according to our Model 1 simulations. In the left plot the red dashed area shows the conditions at which there is a deep convective envelope for a given stellar radius and mass as predicted from our {\tt MESA} simulations, while the dotted teal and circled pink areas show instead the parameter space at which the CE survival is possible based on the assumption made in {\tt StarTrack}, for respectively Z$=0.1$ Z$_\odot$ and Z$=0.01$ Z$_\odot$. In the right plot we show the $R_{\text{MAX}}$ prescriptions from Model 2, Model 3, Model 4a, and Model 4b, and the average {\tt StarTrack} $R_{\text{Int}}$ as a function of $M_{\text{ZAMS}}$.

The {\tt StarTrack} average $R_{\text{Int}}$, as shown by the pink line in the right plot from Figure \ref{fig:CE_conv}, follows the following relation:

\begin{equation}
  \begin{aligned}[b]
     &R_{\text{Int}} = 4.488M_{\rm ZAMS}^3\times10^{-3} - 0.881M_{\text{ZAMS}}^2 +73.556M_{\text{ZAMS}}\\
     &- 1070.186
    \end{aligned}
\label{eq:Rint_ZAMS}
\end{equation}

In Model 1, of all the CE events leading to BH-BH mergers, only $\sim$1 per cent are initiated by donors that would have an outer convective envelope according to our {\tt MESA} simulations. From this comparison it is already evident that detailed modelling gives much more restrictive boundaries for CE survival. This means that, if the donor star does indeed need a convective envelope for the binary to survive the CE phase, many of the CE events simulated with our standard prescription that lead a binary to evolve into a BH-BH system, would lead instead merge with the CE. This in turn means that our CE prescription might overestimate the merger rate densities from Table~\ref{tab:MRD} and Table~\ref{tab:MRD_50Msun}.\\
\\
In the right panel of Fig.~\ref{fig:CE_conv} we compare $R_{\rm MAX}$ from the different models with the distribution of $R_{\rm Int}$ in CE events leading to BH-BH mergers. As expected and already illustrated in the plot on the right of Fig. \ref{fig:CE_conv}, the vast majority of the $R_{\rm Int}$ for CE events ($\sim$~97 per cent) falls below our Model 4b maximum radius limit, which explains the absence of any substantial difference for BH-BH merger rate density predictions between Model 1, Model 2, and Model 4b. This number decreases to $\sim$ 83 per cent if we consider only $M_{\rm ZAMS}$ > 50\,M$_\odot$. For this considered mass range, only $\sim$ 1 and 18 per cent of the stars in BH-BH progenitor binaries for respectively Model 3 and Model 4a would be large enough to initiate a CE phase. This explains why the differences in terms of merger rate densities for BH-BH total masses > 50\,M$_\odot$, as shown in Sec.~\ref{subsec:mass_distr}, are sharper than when we consider the whole mass spectrum. From the comparison between Fig. \ref{fig:RMAX} and the right panel on Fig.~\ref{fig:CE_conv}, we also show that all donor stars with $M_{\rm ZAMS}\gtrsim$40\,M$_\odot$ in BH-BH merger progenitors that initiate a CE phase happen to be beyond the HD limit.\\
\\
In Appendix \ref{App:IntRad} histograms of the $R_{\rm Int}$ distributions for CE, stable RLOF and all RLOF events for BH-BH merger progenitors are provided. 

\section{Conclusions}

In our study we analysed how the theoretical uncertainty on the maximum expansion of massive stars ($M_{\rm ZAMS}$ > 18\,M$_\odot$) can affect the evolution of isolated massive binaries and in turn the formation of gravitational-wave sources.\\
\\
We present five different models to describe the maximum stellar radius ($R_{\rm MAX}$) that massive stars reach throughout their lifetime as a function of their initial mass. Model 1 represents our default {\tt StarTrack} prescription, which adopts the \cite{Hurley_2000} analytic formulae for stellar evolution. Model 2 was retrieved from the {METISSE-BoOST} models (see Sec. \ref{subsec:METISSE-BOOST}) from \cite{Agrawal_2020}. Model 3 is based on the the {\tt METISSE-MESA} models (\citealt{Agrawal_2020}, Sec. \ref{subsec:METISSE-MESA}), while Model 4a and Model 4b are based on our sets of evolutionary models computed with {\tt MESA} (Sec. \ref{subsec:our_MESA}). All the models and their respective predictions are shown in Fig.~\ref{fig:RMAX} .\\
\\
Models from 2 to 4b predict $R_{\rm MAX}$ ($M_{\rm ZAMS}$) that is smaller than what is predicted by the analytic formulae from \cite{Hurley_2000} in Model 1, for some masses by even an order of magnitude. These formulae were fit from a set of stellar evolution models \citep{Pols_1998}. These stellar tracks, when used in population studies for gravitational wave sources, must be extrapolated for initial masses higher than 50\,M$_\odot$ and interpolated for every metallicity that was not considered in the original set of models. This leads to artefacts that can alter our predictions of stellar and binary evolution. 
We show in Fig. \ref{fig:RM_Z} and Fig. \ref{fig:RM_Z_MESA} that the maximum radial expansion of a star as a function of metallicity, as simulated by \cite{Hurley_2000} analytic formulae, does not reproduce what is expected by our {\tt MESA} simulations made with Model 4b, which is most apparent for those initial metallicities and masses at which stellar evolution is interpolated and/or extrapolated.\\
\\
Given each of ours models for $R_{\rm MAX}$ ($M_{\rm ZAMS}$), we estimated the merger rate density and BH mass distribution of double compact object mergers, and compared them to the reported LIGO/Virgo/KAGRA values \citep{LIGO_2021,Abbott2021}. We find that in Model 2 and Model 4b there is no significant change in the estimated BH-BH merger rate density from our reference model with the \cite{Hurley_2000} prescription (see Table \ref{tab:MRD}). For Model 3 and Model 4a, we show that the BH-BH merger rate changes at most by a factor of $\sim$ 3. This is not an extreme deviation from our reference model, since the merger rate density of double compact objects was shown to vary by even orders of magnitude between different models and prescriptions \citep{Mandel_2022}.
By studying the merger rate distribution as a function of the total BH-BH merger mass (for redshifts z < 2), as shown in Fig.~\ref{fig:BHBHMass}, we find no significant difference among the models for $M_{\rm tot}$ < 50\,M$_\odot$.\\
\\
Looking at the most massive BH-BH mergers ($M_{\rm tot}$ > 50\,M$_\odot$), Models 3 and Models 4a predict a much lower rate of events compared to our reference model or to Models 2 and 4b (by factors of $\sim$50 and $\sim$15, respectively, see Table \ref{tab:MRD_50Msun}). This is due to the fact that in Model 3 stars with initial masses larger than 45\,M$_\odot$ never expand to become red supergiants, significantly limiting the parameter space for binary interactions. In a similar but less significant effect, in Model 4a a limited radial expansion arises due to the use of MLT++ convection scheme in {\tt MESA} (as discussed in \citealt{Chun_2018,Klencki_2020}). This is of the utmost importance since radial expansion is needed to initiate Roche lobe overflow events and in turn producing merging double compact objects in the isolated binary evolution channel (with the exception of the chemically homogeneous evolution sub-channel, which is not the subject of our study). We also showed that in any model we presented the most massive binary BH mergers tend to form from common envelope interactions initiated by donors from above the Humphrey-Davidson limit. Since the evolution of massive stars beyond this threshold is still uncertain, we cannot yet know if they further expand under these conditions. We cannot therefore use this fact to test the goodness of our models, but on the other hand we argue that if stars didn't actually expand beyond the Humphrey-Davidson limit, our models would not be able to reproduce the bulk of BH-BH mergers that the current gravitational wave telescopes detect.
\\
\\We also conclude that despite the adoption of \cite{Hurley_2000} formulae in population studies of gravitational wave sources potentially leads to an overestimation of the maximum expansion of massive stars, this does not significantly alter the estimates of merger rate densities in respect to modern detailed evolutionary calculations of stellar radial expansion. This is due to the fact that the vast majority of binary interactions, in the context of the models explored here, occur before stars are able to expand close to their (theoretically uncertain) maximum radius (see Sec. \ref{sec:RLOF_env}). BH-BH merger total mass distribution for low masses ($M_{\rm tot}$ < 50\,M$_\odot$) does not show any difference between the results that employ Hurley's radial expansion and the expansion adopted from various current detailed stellar evolution models. 
For high total BH-BH merger masses ($M_{\rm tot}$ > 50\,M$_\odot$) stellar models of massive stars with modest expansion and most reasonable input physics (Models 2 and 4b) also predict very similar mass distributions to the predictions based on Hurley's prescriptions (Model 1). However, models with more restrictive stellar expansion (Models 3 and 4a) are significantly different (they predict less massive BH-BH mergers) than other updated detailed evolutionary models (Models 2 and 4b) or models based on \cite{Hurley_2000} prescriptions (Model 1). While there are other factors that can alter the BH-BH merger mass distribution (e.g. \citealt{Stevenson_2019,Belczynski_2020_PPSN,Belczynski_2020b,Mapelli_2020,Vink_2021,Belczynski2022,Belczynski_2022b,Briel_2022,Fryer_2022,vanSon_2022}), we suggest that the study of BH-BH mergers beyond a total mass of 50\,M$_\odot$ could help to better constrain the radial evolution of massive stars. At the same time, our results illustrate that the understanding of radial expansion of the most massive stars and the origin of Humphreys-Davidson limit \citep{Humphreys1979,Davies2018,Gilkis_2021,Sabhahit_2022} are of crucial importance for the formation of massive BH-BH mergers via common envelope evolution.\\
\\
Finally, we show in Fig. \ref{fig:CE_conv} that, when compared to detailed evolutionary models, our standard prescription predicts mostly common envelope events during which the donor star in BH-BH merger progenitor binary possesses an outer radiative envelope. This means that, according to \cite{Klencki_2020,Klencki_2021} and \cite{Marchant_2021}, our estimates for common envelope survival and in turn BH-BH mergers could be overestimated. This conclusion is further strengthened by the results from \cite{Gallegos_2021}, which show a significant overestimation of CE survival rates in population synthesis models if compared with detailed ones.

\section*{Acknowledgements}

AR and KB acknowledges support from the Polish National Science Center (NCN) grant Maestro (2018/30/A/ST9/00050). Special thanks go to tens of thousands of citizen-science project "Universe@home" (universeathome.pl) enthusiasts that help to develop the StarTrack population synthesis code used in this study. AR and KB thank J. Andrews for the insight about the metallicity-dependent radial evolution of stars in population synthesis codes. AR and KB acknowledge the help of A. Olejak and R. Smolec for their helpful feedback. TS acknowledges support from the European Union's Horizon 2020 under the Marie Skłodowska-Curie grant agreement No 101024605. JK acknowledges support from an ESO Fellowship. This research was funded in part by the National Science Center (NCN), Poland under grant number OPUS 2021/41/B/ST9/00757. For the purpose of Open Access, the author has applied a CC-BY public copyright license to any Author Accepted Manuscript (AAM) version arising from this submission.

\section*{Data Availability}

The data presented in this work can be made available based on the individual request to the corresponding authors.



\bibliographystyle{mnras}
\bibliography{example} 

\begin{thebibliography}{}
\makeatletter
\relax
\def\mn@urlcharsother{\let\do\@makeother \do\$\do\&\do\#\do\^\do\_\do\%\do\~}
\def\mn@doi{\begingroup\mn@urlcharsother \@ifnextchar [ {\mn@doi@}
  {\mn@doi@[]}}
\def\mn@doi@[#1]#2{\def\@tempa{#1}\ifx\@tempa\@empty \href
  {http://dx.doi.org/#2} {doi:#2}\else \href {http://dx.doi.org/#2} {#1}\fi
  \endgroup}
\def\mn@eprint#1#2{\mn@eprint@#1:#2::\@nil}
\def\mn@eprint@arXiv#1{\href {http://arxiv.org/abs/#1} {{\tt arXiv:#1}}}
\def\mn@eprint@dblp#1{\href {http://dblp.uni-trier.de/rec/bibtex/#1.xml}
  {dblp:#1}}
\def\mn@eprint@#1:#2:#3:#4\@nil{\def\@tempa {#1}\def\@tempb {#2}\def\@tempc
  {#3}\ifx \@tempc \@empty \let \@tempc \@tempb \let \@tempb \@tempa \fi \ifx
  \@tempb \@empty \def\@tempb {arXiv}\fi \@ifundefined
  {mn@eprint@\@tempb}{\@tempb:\@tempc}{\expandafter \expandafter \csname
  mn@eprint@\@tempb\endcsname \expandafter{\@tempc}}}

\bibitem[\protect\citeauthoryear{Abbott et~al.,}{Abbott
  et~al.}{2021a}]{LIGO_2021}
Abbott B.~P.,  et~al., 2021a, The population of merging compact binaries
  inferred using gravitational waves through GWTC-3 (\mn@eprint {arXiv}
  {2111.03634})

\bibitem[\protect\citeauthoryear{Abbott et~al.,}{Abbott
  et~al.}{2021b}]{Abbott2021}
Abbott R.,  et~al., 2021b, \mn@doi [The Astrophysical Journal Letters]
  {10.3847/2041-8213/abe949}, 913, L7

\bibitem[\protect\citeauthoryear{Agrawal, Hurley, Stevenson, Szécsi  \&
  Flynn}{Agrawal et~al.}{2020}]{Agrawal_2020}
Agrawal P.,  Hurley J.,  Stevenson S.,  Szécsi D.,   Flynn C.,  2020, \mn@doi
  [Monthly Notices of the Royal Astronomical Society] {10.1093/mnras/staa2264},
  497, 4549–4564

\bibitem[\protect\citeauthoryear{{Agrawal}, {Stevenson}, {Sz{\'e}csi}  \&
  {Hurley}}{{Agrawal} et~al.}{2021}]{Agrawal:2021b}
{Agrawal} P.,  {Stevenson} S.,  {Sz{\'e}csi} D.,   {Hurley} J.,  2021, arXiv
  e-prints, \href {https://ui.adsabs.harvard.edu/abs/2021arXiv211202801A} {p.
  arXiv:2112.02801}

\bibitem[\protect\citeauthoryear{{Agrawal}, {Sz{\'e}csi}, {Stevenson},
  {Eldridge}  \& {Hurley}}{{Agrawal} et~al.}{2022}]{Agrawal:2022}
{Agrawal} P.,  {Sz{\'e}csi} D.,  {Stevenson} S.,  {Eldridge} J.~J.,   {Hurley}
  J.,  2022, \mn@doi [\mnras] {10.1093/mnras/stac930}, \href
  {https://ui.adsabs.harvard.edu/abs/2022MNRAS.512.5717A} {512, 5717}

\bibitem[\protect\citeauthoryear{Ali-Ha\"{\i}moud, Kovetz  \&
  Kamionkowski}{Ali-Ha\"{\i}moud et~al.}{2017}]{Haimoud2017}
Ali-Ha\"{\i}moud Y.,  Kovetz E.~D.,   Kamionkowski M.,  2017, \mn@doi [Phys.
  Rev. D] {10.1103/PhysRevD.96.123523}, 96, 123523

\bibitem[\protect\citeauthoryear{Antonini \& Rasio}{Antonini \&
  Rasio}{2016}]{Antonini_2016}
Antonini F.,  Rasio F.~A.,  2016, \mn@doi [The Astrophysical Journal]
  {10.3847/0004-637x/831/2/187}, 831, 187

\bibitem[\protect\citeauthoryear{Bartos, Kocsis, Haiman  \& M{\'{a}}rka}{Bartos
  et~al.}{2017}]{Bartos_2017}
Bartos I.,  Kocsis B.,  Haiman Z.,   M{\'{a}}rka S.,  2017, \mn@doi [The
  Astrophysical Journal] {10.3847/1538-4357/835/2/165}, 835, 165

\bibitem[\protect\citeauthoryear{Belczynski}{Belczynski}{2020}]{Belczynski_2020_PPSN}
Belczynski K.,  2020, \mn@doi [The Astrophysical Journal]
  {10.3847/2041-8213/abcbf1}, 905, L15

\bibitem[\protect\citeauthoryear{Belczynski, Kalogera  \& Bulik}{Belczynski
  et~al.}{2002}]{Belczynski_2002}
Belczynski K.,  Kalogera V.,   Bulik T.,  2002, \mn@doi [The Astrophysical
  Journal] {10.1086/340304}, 572, 407

\bibitem[\protect\citeauthoryear{Belczynski, Taam, Kalogera, Rasio  \&
  Bulik}{Belczynski et~al.}{2007}]{Belczynski_2007}
Belczynski K.,  Taam R.~E.,  Kalogera V.,  Rasio F.~A.,   Bulik T.,  2007,
  \mn@doi [The Astrophysical Journal] {10.1086/513562}, 662, 504

\bibitem[\protect\citeauthoryear{Belczynski, Kalogera, Rasio, Taam, Zezas,
  Bulik, Maccarone  \& Ivanova}{Belczynski et~al.}{2008}]{Belczynski_2008}
Belczynski K.,  Kalogera V.,  Rasio F.~A.,  Taam R.~E.,  Zezas A.,  Bulik T.,
  Maccarone T.~J.,   Ivanova N.,  2008, \mn@doi [The Astrophysical Journal
  Supplement Series] {10.1086/521026}, 174, 223

\bibitem[\protect\citeauthoryear{Belczynski, Bulik, Fryer, Ruiter, Valsecchi,
  Vink  \& Hurley}{Belczynski et~al.}{2010}]{Belczynski_2010b}
Belczynski K.,  Bulik T.,  Fryer C.~L.,  Ruiter A.,  Valsecchi F.,  Vink J.~S.,
    Hurley J.~R.,  2010, \mn@doi [The Astrophysical Journal]
  {10.1088/0004-637X/714/2/1217}, 714, 1217

\bibitem[\protect\citeauthoryear{Belczynski, Wiktorowicz, Fryer, Holz  \&
  Kalogera}{Belczynski et~al.}{2012}]{Belczynski_2012}
Belczynski K.,  Wiktorowicz G.,  Fryer C.~L.,  Holz D.~E.,   Kalogera V.,
  2012, \mn@doi [The Astrophysical Journal] {10.1088/0004-637x/757/1/91}, 757,
  91

\bibitem[\protect\citeauthoryear{{Belczynski} et~al.,}{{Belczynski}
  et~al.}{2016}]{Belczynski_2016}
{Belczynski} et~al., 2016, \mn@doi [A\&A] {10.1051/0004-6361/201628980}, 594,
  A97

\bibitem[\protect\citeauthoryear{{Belczynski} et~al.,}{{Belczynski}
  et~al.}{2020a}]{Belczynski_2020}
{Belczynski} et~al., 2020a, \mn@doi [A\&A] {10.1051/0004-6361/201936528}, 636,
  A104

\bibitem[\protect\citeauthoryear{{Belczynski} et~al.,}{{Belczynski}
  et~al.}{2020b}]{Belczynski_2020b}
{Belczynski} K.,  et~al., 2020b, \mn@doi [\apj] {10.3847/1538-4357/ab6d77},
  \href {https://ui.adsabs.harvard.edu/abs/2020ApJ...890..113B} {890, 113}

\bibitem[\protect\citeauthoryear{Belczynski et~al.,}{Belczynski
  et~al.}{2022a}]{Belczynski2022}
Belczynski K.,  et~al., 2022a, \mn@doi [The Astrophysical Journal]
  {10.3847/1538-4357/ac375a}, 925, 69

\bibitem[\protect\citeauthoryear{{Belczynski}, {Doctor}, {Zevin}, {Olejak},
  {Banerje}  \& {Chattopadhyay}}{{Belczynski} et~al.}{2022b}]{Belczynski_2022b}
{Belczynski} K.,  {Doctor} Z.,  {Zevin} M.,  {Olejak} A.,  {Banerje} S.,
  {Chattopadhyay} D.,  2022b, \mn@doi [\apj] {10.3847/1538-4357/ac8167}, \href
  {https://ui.adsabs.harvard.edu/abs/2022ApJ...935..126B} {935, 126}

\bibitem[\protect\citeauthoryear{Bellovary, Low, McKernan  \& Ford}{Bellovary
  et~al.}{2016}]{Bellovary_2016}
Bellovary J.~M.,  Low M.-M.~M.,  McKernan B.,   Ford K. E.~S.,  2016, \mn@doi
  [The Astrophysical Journal] {10.3847/2041-8205/819/2/l17}, 819, L17

\bibitem[\protect\citeauthoryear{Bird, Cholis, Mu\~noz, Ali-Ha\"{\i}moud,
  Kamionkowski, Kovetz, Raccanelli  \& Riess}{Bird et~al.}{2016}]{Bird2016}
Bird S.,  Cholis I.,  Mu\~noz J.~B.,  Ali-Ha\"{\i}moud Y.,  Kamionkowski M.,
  Kovetz E.~D.,  Raccanelli A.,   Riess A.~G.,  2016, \mn@doi [Phys. Rev.
  Lett.] {10.1103/PhysRevLett.116.201301}, 116, 201301

\bibitem[\protect\citeauthoryear{{Boffin} \& {Jorissen}}{{Boffin} \&
  {Jorissen}}{1988}]{BoffinJorissen1988}
{Boffin} H.~M.~J.,  {Jorissen} A.,  1988, \aap, \href
  {https://ui.adsabs.harvard.edu/abs/1988A&A...205..155B} {205, 155}

\bibitem[\protect\citeauthoryear{{B{\"o}hm-Vitense}}{{B{\"o}hm-Vitense}}{1958}]{Bohm-Vitense_1958}
{B{\"o}hm-Vitense} E.,  1958, \zap, \href
  {https://ui.adsabs.harvard.edu/abs/1958ZA.....46..108B} {46, 108}

\bibitem[\protect\citeauthoryear{{Bonaca} et~al.,}{{Bonaca}
  et~al.}{2012}]{Bonaca_2012}
{Bonaca} A.,  et~al., 2012, \mn@doi [\apjl] {10.1088/2041-8205/755/1/L12},
  \href {https://ui.adsabs.harvard.edu/abs/2012ApJ...755L..12B} {755, L12}

\bibitem[\protect\citeauthoryear{{Bondi} \& {Hoyle}}{{Bondi} \&
  {Hoyle}}{1944}]{BondiHoyle1944}
{Bondi} H.,  {Hoyle} F.,  1944, \mn@doi [\mnras] {10.1093/mnras/104.5.273},
  \href {https://ui.adsabs.harvard.edu/abs/1944MNRAS.104..273B} {104, 273}

\bibitem[\protect\citeauthoryear{{Briel}, {Stevance}  \& {Eldridge}}{{Briel}
  et~al.}{2022}]{Briel_2022}
{Briel} M.~M.,  {Stevance} H.~F.,   {Eldridge} J.~J.,  2022, arXiv e-prints,
  \href {https://ui.adsabs.harvard.edu/abs/2022arXiv220613842B} {p.
  arXiv:2206.13842}

\bibitem[\protect\citeauthoryear{{Brott} et~al.,}{{Brott}
  et~al.}{2011}]{Brott_2011}
{Brott} I.,  et~al., 2011, \mn@doi [\aap] {10.1051/0004-6361/201016113}, \href
  {http://adsabs.harvard.edu/abs/2011A%26A...530A.115B} {530, A115}

\bibitem[\protect\citeauthoryear{{Burrows}, {Hubbard}, {Saumon}  \&
  {Lunine}}{{Burrows} et~al.}{1993}]{Burrows_1993}
{Burrows} A.,  {Hubbard} W.~B.,  {Saumon} D.,   {Lunine} J.~I.,  1993, \mn@doi
  [\apj] {10.1086/172427}, \href
  {https://ui.adsabs.harvard.edu/abs/1993ApJ...406..158B} {406, 158}

\bibitem[\protect\citeauthoryear{{Casares}}{{Casares}}{2007}]{Casares_2007}
{Casares} J.,  2007, ] {10.1017/S1743921307004590}, \href
  {https://ui.adsabs.harvard.edu/abs/2007IAUS..238....3C} {238, 3}

\bibitem[\protect\citeauthoryear{Chen \& Huang}{Chen \&
  Huang}{2018}]{Chen_2018}
Chen Z.-C.,  Huang Q.-G.,  2018, \mn@doi [The Astrophysical Journal]
  {10.3847/1538-4357/aad6e2}, 864, 61

\bibitem[\protect\citeauthoryear{{Choi}, {Dotter}, {Conroy}, {Cantiello},
  {Paxton}  \& {Johnson}}{{Choi} et~al.}{2016}]{Choi:2016MIST}
{Choi} J.,  {Dotter} A.,  {Conroy} C.,  {Cantiello} M.,  {Paxton} B.,
  {Johnson} B.~D.,  2016, \mn@doi [\apj] {10.3847/0004-637X/823/2/102}, \href
  {http://adsabs.harvard.edu/abs/2016ApJ...823..102C} {823, 102}

\bibitem[\protect\citeauthoryear{{Chun}, {Yoon}, {Jung}, {Kim}  \&
  {Kim}}{{Chun} et~al.}{2018}]{Chun_2018}
{Chun} S.-H.,  {Yoon} S.-C.,  {Jung} M.-K.,  {Kim} D.~U.,   {Kim} J.,  2018,
  \mn@doi [\apj] {10.3847/1538-4357/aa9a37}, \href
  {https://ui.adsabs.harvard.edu/abs/2018ApJ...853...79C} {853, 79}

\bibitem[\protect\citeauthoryear{{Davies}, {Crowther}  \& {Beasor}}{{Davies}
  et~al.}{2018}]{Davies2018}
{Davies} B.,  {Crowther} P.~A.,   {Beasor} E.~R.,  2018, \mn@doi [\mnras]
  {10.1093/mnras/sty1302}, \href
  {https://ui.adsabs.harvard.edu/abs/2018MNRAS.478.3138D} {478, 3138}

\bibitem[\protect\citeauthoryear{{Dominik}, {Belczynski}, {Fryer}, {Holz},
  {Berti}, {Bulik}, {Mandel}  \& {O'Shaughnessy}}{{Dominik}
  et~al.}{2012}]{Dominik_2012}
{Dominik} M.,  {Belczynski} K.,  {Fryer} C.,  {Holz} D.~E.,  {Berti} E.,
  {Bulik} T.,  {Mandel} I.,   {O'Shaughnessy} R.,  2012, \mn@doi [\apj]
  {10.1088/0004-637X/759/1/52}, \href
  {https://ui.adsabs.harvard.edu/abs/2012ApJ...759...52D} {759, 52}

\bibitem[\protect\citeauthoryear{Dorn-Wallenstein, Levesque, Neugent,
  Davenport, Morris  \& Gootkin}{Dorn-Wallenstein
  et~al.}{2020}]{Dorn_Wallenstein_2020}
Dorn-Wallenstein T.~Z.,  Levesque E.~M.,  Neugent K.~F.,  Davenport J. R.~A.,
  Morris B.~M.,   Gootkin K.,  2020, \mn@doi [The Astrophysical Journal]
  {10.3847/1538-4357/abb318}, 902, 24

\bibitem[\protect\citeauthoryear{{Dufton}, {McErlean}, {Lennon}  \&
  {Ryans}}{{Dufton} et~al.}{2000}]{Dufton2000}
{Dufton} P.~L.,  {McErlean} N.~D.,  {Lennon} D.~J.,   {Ryans} R.~S.~I.,  2000,
  \aap, \href {https://ui.adsabs.harvard.edu/abs/2000A&A...353..311D} {353,
  311}

\bibitem[\protect\citeauthoryear{{Evans} \& {Howarth}}{{Evans} \&
  {Howarth}}{2003}]{Evans2003}
{Evans} C.~J.,  {Howarth} I.~D.,  2003, \mn@doi [\mnras]
  {10.1046/j.1365-2966.2003.07038.x}, \href
  {https://ui.adsabs.harvard.edu/abs/2003MNRAS.345.1223E} {345, 1223}

\bibitem[\protect\citeauthoryear{Farrell, Groh, Meynet  \& Eldridge}{Farrell
  et~al.}{2021}]{farrell2021}
Farrell E.,  Groh J.,  Meynet G.,   Eldridge J.,  2021, Understanding the
  evolution of massive stars (\mn@eprint {arXiv} {2109.02488})

\bibitem[\protect\citeauthoryear{{Fryer}, {Woosley}  \& {Hartmann}}{{Fryer}
  et~al.}{1999}]{FWH1999}
{Fryer} C.~L.,  {Woosley} S.~E.,   {Hartmann} D.~H.,  1999, \mn@doi [\apj]
  {10.1086/307992}, \href
  {https://ui.adsabs.harvard.edu/abs/1999ApJ...526..152F} {526, 152}

\bibitem[\protect\citeauthoryear{Fryer, Belczynski, Wiktorowicz, Dominik,
  Kalogera  \& Holz}{Fryer et~al.}{2012}]{Fryer_2012}
Fryer C.~L.,  Belczynski K.,  Wiktorowicz G.,  Dominik M.,  Kalogera V.,   Holz
  D.~E.,  2012, \mn@doi [The Astrophysical Journal]
  {10.1088/0004-637x/749/1/91}, 749, 91

\bibitem[\protect\citeauthoryear{{Fryer}, {Olejak}  \& {Belczynski}}{{Fryer}
  et~al.}{2022}]{Fryer_2022}
{Fryer} C.~L.,  {Olejak} A.,   {Belczynski} K.,  2022, \mn@doi [\apj]
  {10.3847/1538-4357/ac6ac9}, \href
  {https://ui.adsabs.harvard.edu/abs/2022ApJ...931...94F} {931, 94}

\bibitem[\protect\citeauthoryear{{Gaia Collaboration}}{{Gaia
  Collaboration}}{2018}]{GaiaDR2}
{Gaia Collaboration} 2018, VizieR Online Data Catalog, \href
  {https://ui.adsabs.harvard.edu/abs/2018yCat.1345....0G} {p. I/345}

\bibitem[\protect\citeauthoryear{{Gallegos-Garcia}, {Berry}, {Marchant}  \&
  {Kalogera}}{{Gallegos-Garcia} et~al.}{2021}]{Gallegos_2021}
{Gallegos-Garcia} M.,  {Berry} C. P.~L.,  {Marchant} P.,   {Kalogera} V.,
  2021, \mn@doi [\apj] {10.3847/1538-4357/ac2610}, \href
  {https://ui.adsabs.harvard.edu/abs/2021ApJ...922..110G} {922, 110}

\bibitem[\protect\citeauthoryear{{Georgy, C.}, {Ekstr\"om, S.}, {Granada, A.},
  {Meynet, G.}, {Mowlavi, N.}, {Eggenberger, P.}  \& {Maeder, A.}}{{Georgy, C.}
  et~al.}{2013}]{Georgy_2013}
{Georgy, C.} {Ekstr\"om, S.} {Granada, A.} {Meynet, G.} {Mowlavi, N.}
  {Eggenberger, P.}  {Maeder, A.} 2013, \mn@doi [A\&A]
  {10.1051/0004-6361/201220558}, 553, A24

\bibitem[\protect\citeauthoryear{{Gilkis}, {Shenar}, {Ramachandran}, {Jermyn},
  {Mahy}, {Oskinova}, {Arcavi}  \& {Sana}}{{Gilkis} et~al.}{2021}]{Gilkis_2021}
{Gilkis} A.,  {Shenar} T.,  {Ramachandran} V.,  {Jermyn} A.~S.,  {Mahy} L.,
  {Oskinova} L.~M.,  {Arcavi} I.,   {Sana} H.,  2021, \mn@doi [\mnras]
  {10.1093/mnras/stab383}, \href
  {https://ui.adsabs.harvard.edu/abs/2021MNRAS.503.1884G} {503, 1884}

\bibitem[\protect\citeauthoryear{{Glebbeek}, {Gaburov}, {de Mink}, {Pols}  \&
  {Portegies Zwart}}{{Glebbeek} et~al.}{2009}]{Glebbeek:2009}
{Glebbeek} E.,  {Gaburov} E.,  {de Mink} S.~E.,  {Pols} O.~R.,   {Portegies
  Zwart} S.~F.,  2009, \mn@doi [\aap] {10.1051/0004-6361/200810425}, \href
  {https://ui.adsabs.harvard.edu/abs/2009A&A...497..255G} {497, 255}

\bibitem[\protect\citeauthoryear{{Grevesse} \& {Sauval}}{{Grevesse} \&
  {Sauval}}{1998}]{Grevesse&Sauval_1998}
{Grevesse} N.,  {Sauval} A.~J.,  1998, \mn@doi [\ssr]
  {10.1023/A:1005161325181}, \href
  {https://ui.adsabs.harvard.edu/abs/1998SSRv...85..161G} {85, 161}

\bibitem[\protect\citeauthoryear{{Heger}, {Langer}  \& {Woosley}}{{Heger}
  et~al.}{2000}]{Heger_2000}
{Heger} A.,  {Langer} N.,   {Woosley} S.~E.,  2000, \mn@doi [\apj]
  {10.1086/308158}, \href
  {https://ui.adsabs.harvard.edu/abs/2000ApJ...528..368H} {528, 368}

\bibitem[\protect\citeauthoryear{{Heger}, {Fryer}, {Woosley}, {Langer}  \&
  {Hartmann}}{{Heger} et~al.}{2003}]{Heger_2003}
{Heger} A.,  {Fryer} C.~L.,  {Woosley} S.~E.,  {Langer} N.,   {Hartmann} D.~H.,
   2003, \mn@doi [\apj] {10.1086/375341}, \href
  {https://ui.adsabs.harvard.edu/abs/2003ApJ...591..288H} {591, 288}

\bibitem[\protect\citeauthoryear{{Higgins} \& {Vink}}{{Higgins} \&
  {Vink}}{2019}]{HigginsVink2019}
{Higgins} {Vink} 2019, \mn@doi [A\&A] {10.1051/0004-6361/201834123}, 622, A50

\bibitem[\protect\citeauthoryear{Hobbs, Lorimer, Lyne  \& Kramer}{Hobbs
  et~al.}{2005}]{Hobbs_2005}
Hobbs G.,  Lorimer D.~R.,  Lyne A.~G.,   Kramer M.,  2005, \mn@doi [Monthly
  Notices of the Royal Astronomical Society]
  {10.1111/j.1365-2966.2005.09087.x}, 360, 974

\bibitem[\protect\citeauthoryear{Horvath, Rocha, Bernardo, de Avellar  \&
  Valentim}{Horvath et~al.}{2020}]{horvath2020}
Horvath J.~E.,  Rocha L.~S.,  Bernardo A. L.~C.,  de Avellar M. G.~B.,
  Valentim R.,  2020, Birth events, masses and the maximum mass of Compact
  Stars (\mn@eprint {arXiv} {2011.08157})

\bibitem[\protect\citeauthoryear{{Humphreys} \& {Davidson}}{{Humphreys} \&
  {Davidson}}{1979}]{Humphreys1979}
{Humphreys} R.~M.,  {Davidson} K.,  1979, \mn@doi [\apj] {10.1086/157301},
  \href {https://ui.adsabs.harvard.edu/abs/1979ApJ...232..409H} {232, 409}

\bibitem[\protect\citeauthoryear{Humphreys \& Davidson}{Humphreys \&
  Davidson}{1994}]{Humphreys_1994}
Humphreys R.~M.,  Davidson K.,  1994, \mn@doi [Publications of the Astronomical
  Society of the Pacific] {10.1086/133478}, 106, 1025

\bibitem[\protect\citeauthoryear{Hurley, Pols  \& Tout}{Hurley
  et~al.}{2000}]{Hurley_2000}
Hurley J.~R.,  Pols O.~R.,   Tout C.~A.,  2000, \mn@doi [Monthly Notices of the
  Royal Astronomical Society] {10.1046/j.1365-8711.2000.03426.x}, 315, 543

\bibitem[\protect\citeauthoryear{Kesseli et~al.,}{Kesseli
  et~al.}{2019}]{Kesseli_2019}
Kesseli A.~Y.,  et~al., 2019, \mn@doi [The Astronomical Journal]
  {10.3847/1538-3881/aae982}, 157, 63

\bibitem[\protect\citeauthoryear{{King}, {Davies}, {Ward}, {Fabbiano}  \&
  {Elvis}}{{King} et~al.}{2001}]{King2001}
{King} A.~R.,  {Davies} M.~B.,  {Ward} M.~J.,  {Fabbiano} G.,   {Elvis} M.,
  2001, \mn@doi [\apjl] {10.1086/320343}, \href
  {https://ui.adsabs.harvard.edu/abs/2001ApJ...552L.109K} {552, L109}

\bibitem[\protect\citeauthoryear{{Klencki}, {Nelemans}, {Istrate}  \&
  {Pols}}{{Klencki} et~al.}{2020}]{Klencki_2020}
{Klencki} J.,  {Nelemans} G.,  {Istrate} A.~G.,   {Pols} O.,  2020, \mn@doi
  [\aap] {10.1051/0004-6361/202037694}, \href
  {https://ui.adsabs.harvard.edu/abs/2020A&A...638A..55K} {638, A55}

\bibitem[\protect\citeauthoryear{{Klencki}, {Nelemans, Gijs}, {Istrate, Alina
  G.}  \& {Chruslinska, Martyna}}{{Klencki} et~al.}{2021}]{Klencki_2021}
{Klencki} {Nelemans, Gijs} {Istrate, Alina G.}  {Chruslinska, Martyna} 2021,
  \mn@doi [A\&A] {10.1051/0004-6361/202038707}, 645, A54

\bibitem[\protect\citeauthoryear{{Kozai}}{{Kozai}}{1962}]{Kozai_1962}
{Kozai} Y.,  1962, \mn@doi [\aj] {10.1086/108790}, \href
  {https://ui.adsabs.harvard.edu/abs/1962AJ.....67..591K} {67, 591}

\bibitem[\protect\citeauthoryear{{Kroupa}}{{Kroupa}}{2002}]{Kroupa_2002}
{Kroupa} P.,  2002, \mn@doi [Science] {10.1126/science.1067524}, \href
  {https://ui.adsabs.harvard.edu/abs/2002Sci...295...82K} {295, 82}

\bibitem[\protect\citeauthoryear{{Kroupa}, {Tout}  \& {Gilmore}}{{Kroupa}
  et~al.}{1993}]{Kroupa_1993}
{Kroupa} P.,  {Tout} C.~A.,   {Gilmore} G.,  1993, \mn@doi [\mnras]
  {10.1093/mnras/262.3.545}, \href
  {https://ui.adsabs.harvard.edu/abs/1993MNRAS.262..545K} {262, 545}

\bibitem[\protect\citeauthoryear{{Laplace}, {G\"otberg}, {de Mink}, {Justham}
  \& {Farmer}}{{Laplace} et~al.}{2020}]{Laplace_2020}
{Laplace} E.,  {G\"otberg} Y.,  {de Mink} S.~E.,  {Justham} S.,   {Farmer} R.,
  2020, \mn@doi [A\&A] {10.1051/0004-6361/201937300}, 637, A6

\bibitem[\protect\citeauthoryear{{Ledoux}}{{Ledoux}}{1947}]{Ledoux_1947}
{Ledoux} P.,  1947, \mn@doi [\apj] {10.1086/144905}, \href
  {https://ui.adsabs.harvard.edu/abs/1947ApJ...105..305L} {105, 305}

\bibitem[\protect\citeauthoryear{{Levesque}, {Massey}, {Olsen}, {Plez},
  {Josselin}, {Maeder}  \& {Meynet}}{{Levesque} et~al.}{2005}]{Levesque2005}
{Levesque} E.~M.,  {Massey} P.,  {Olsen} K.~A.~G.,  {Plez} B.,  {Josselin} E.,
  {Maeder} A.,   {Meynet} G.,  2005, \mn@doi [\apj] {10.1086/430901}, \href
  {https://ui.adsabs.harvard.edu/abs/2005ApJ...628..973L} {628, 973}

\bibitem[\protect\citeauthoryear{Lidov}{Lidov}{1962}]{Lidov_1962}
Lidov M.,  1962, \mn@doi [Planetary and Space Science]
  {https://doi.org/10.1016/0032-0633(62)90129-0}, 9, 719

\bibitem[\protect\citeauthoryear{Liu \& Lai}{Liu \& Lai}{2018}]{Liu_2018}
Liu B.,  Lai D.,  2018, \mn@doi [The Astrophysical Journal]
  {10.3847/1538-4357/aad09f}, 863, 68

\bibitem[\protect\citeauthoryear{Maeder \& Maeynet}{Maeder \&
  Maeynet}{2000}]{MaederMeynet2000}
Maeder A.,  Maeynet G.,  2000, \mn@doi [Annual Review of Astronomy and
  Astrophysics] {10.1146/annurev.astro.38.1.143}, 38, 143

\bibitem[\protect\citeauthoryear{Mandel \& Broekgaarden}{Mandel \&
  Broekgaarden}{2021}]{mandel2021rates}
Mandel I.,  Broekgaarden F.~S.,  2021, Rates of Compact Object Coalescences
  (\mn@eprint {arXiv} {2107.14239})

\bibitem[\protect\citeauthoryear{Mandel \& Broekgaarden}{Mandel \&
  Broekgaarden}{2022}]{Mandel_2022}
Mandel I.,  Broekgaarden F.~S.,  2022, \mn@doi [Living Reviews in Relativity]
  {10.1007/s41114-021-00034-3}, 25

\bibitem[\protect\citeauthoryear{Mandel \& de Mink}{Mandel \&
  de Mink}{2016}]{MandeldeMink_2016}
Mandel I.,  de Mink S.~E.,  2016, \mn@doi [Monthly Notices of the Royal
  Astronomical Society] {10.1093/mnras/stw379}, 458, 2634

\bibitem[\protect\citeauthoryear{{Mapelli}, {Spera}, {Montanari}, {Limongi},
  {Chieffi}, {Giacobbo}, {Bressan}  \& {Bouffanais}}{{Mapelli}
  et~al.}{2020}]{Mapelli_2020}
{Mapelli} M.,  {Spera} M.,  {Montanari} E.,  {Limongi} M.,  {Chieffi} A.,
  {Giacobbo} N.,  {Bressan} A.,   {Bouffanais} Y.,  2020, \mn@doi [\apj]
  {10.3847/1538-4357/ab584d}, \href
  {https://ui.adsabs.harvard.edu/abs/2020ApJ...888...76M} {888, 76}

\bibitem[\protect\citeauthoryear{{Marchant, Pablo}, {Langer, Norbert},
  {Podsiadlowski, Philipp}, {Tauris, Thomas M.}  \& {Moriya, Takashi
  J.}}{{Marchant, Pablo} et~al.}{2016}]{Marchant_2016}
{Marchant, Pablo} {Langer, Norbert} {Podsiadlowski, Philipp} {Tauris, Thomas
  M.}  {Moriya, Takashi J.} 2016, \mn@doi [A\&A] {10.1051/0004-6361/201628133},
  588, A50

\bibitem[\protect\citeauthoryear{Marchant, Pappas, Gallegos-Garcia, Berry,
  Taam, Kalogera  \& Podsiadlowski}{Marchant et~al.}{2021}]{Marchant_2021}
Marchant P.,  Pappas K. M.~W.,  Gallegos-Garcia M.,  Berry C. P.~L.,  Taam
  R.~E.,  Kalogera V.,   Podsiadlowski P.,  2021, \mn@doi [Astronomy {\&}
  Astrophysics] {10.1051/0004-6361/202039992}

\bibitem[\protect\citeauthoryear{McKernan et~al.,}{McKernan
  et~al.}{2018}]{McKernan_2018}
McKernan B.,  et~al., 2018, \mn@doi [The Astrophysical Journal]
  {10.3847/1538-4357/aadae5}, 866, 66

\bibitem[\protect\citeauthoryear{{Mennekens} \& {Vanbeveren}}{{Mennekens} \&
  {Vanbeveren}}{2014}]{Mennekens_2014}
{Mennekens} {Vanbeveren} 2014, \mn@doi [A\&A] {10.1051/0004-6361/201322198},
  564, A134

\bibitem[\protect\citeauthoryear{Miller \& Lauburg}{Miller \&
  Lauburg}{2009}]{Miller_2009}
Miller M.~C.,  Lauburg V.~M.,  2009, \mn@doi [The Astrophysical Journal]
  {10.1088/0004-637x/692/1/917}, 692, 917

\bibitem[\protect\citeauthoryear{{Moe} \& {Di Stefano}}{{Moe} \& {Di
  Stefano}}{2017}]{Moe2017}
{Moe} M.,  {Di Stefano} R.,  2017, \mn@doi [\apjs] {10.3847/1538-4365/aa6fb6},
  \href {https://ui.adsabs.harvard.edu/abs/2017ApJS..230...15M} {230, 15}

\bibitem[\protect\citeauthoryear{Müller, Heger, Liptai  \& Cameron}{Müller
  et~al.}{2016}]{Muller_2016}
Müller B.,  Heger A.,  Liptai D.,   Cameron J.~B.,  2016, \mn@doi [Monthly
  Notices of the Royal Astronomical Society] {10.1093/mnras/stw1083}, 460, 742

\bibitem[\protect\citeauthoryear{{Neugent}, {Massey}, {Skiff}, {Drout},
  {Meynet}  \& {Olsen}}{{Neugent} et~al.}{2010}]{Neugent2010}
{Neugent} K.~F.,  {Massey} P.,  {Skiff} B.,  {Drout} M.~R.,  {Meynet} G.,
  {Olsen} K. A.~G.,  2010, \mn@doi [\apj] {10.1088/0004-637X/719/2/1784}, \href
  {https://ui.adsabs.harvard.edu/abs/2010ApJ...719.1784N} {719, 1784}

\bibitem[\protect\citeauthoryear{{Ng}, {Vitale}, {Farr}  \& {Rodriguez}}{{Ng}
  et~al.}{2021}]{Ng_2021}
{Ng} K. K.~Y.,  {Vitale} S.,  {Farr} W.~M.,   {Rodriguez} C.~L.,  2021, \mn@doi
  [\apjl] {10.3847/2041-8213/abf8be}, \href
  {https://ui.adsabs.harvard.edu/abs/2021ApJ...913L...5N} {913, L5}

\bibitem[\protect\citeauthoryear{{Nugis} \& {Lamers}}{{Nugis} \&
  {Lamers}}{2000}]{NugisLamers2000}
{Nugis} T.,  {Lamers} H.~J.~G.~L.~M.,  2000, \aap, \href
  {https://ui.adsabs.harvard.edu/abs/2000A&A...360..227N} {360, 227}

\bibitem[\protect\citeauthoryear{{Olejak}, {Fishbach}, {Belczynski}, {Holz},
  {Lasota}, {Miller}  \& {Bulik}}{{Olejak} et~al.}{2020}]{Olejak_2020}
{Olejak} A.,  {Fishbach} M.,  {Belczynski} K.,  {Holz} D.~E.,  {Lasota} J.~P.,
  {Miller} M.~C.,   {Bulik} T.,  2020, \mn@doi [\apjl]
  {10.3847/2041-8213/abb5b5}, \href
  {https://ui.adsabs.harvard.edu/abs/2020ApJ...901L..39O} {901, L39}

\bibitem[\protect\citeauthoryear{Olejak, Belczynski  \& Ivanova}{Olejak
  et~al.}{2021}]{Olejak_2021}
Olejak A.,  Belczynski K.,   Ivanova N.,  2021, \mn@doi [Astronomy &
  Astrophysics] {10.1051/0004-6361/202140520}, 651, A100

\bibitem[\protect\citeauthoryear{{Orosz}}{{Orosz}}{2003}]{Orosz_2003}
{Orosz} J.~A.,  2003, ] {10.48550/arXiv.astro-ph/0209041}, \href
  {https://ui.adsabs.harvard.edu/abs/2003IAUS..212..365O} {212, 365}

\bibitem[\protect\citeauthoryear{Owocki, Gayley  \& Shaviv}{Owocki
  et~al.}{2004}]{Owocki_2004}
Owocki S.~P.,  Gayley K.~G.,   Shaviv N.~J.,  2004, \mn@doi [The Astrophysical
  Journal] {10.1086/424910}, 616, 525

\bibitem[\protect\citeauthoryear{{Paczynski}}{{Paczynski}}{1976}]{Paczynski_1976}
{Paczynski} B.,  1976, SAO/NASA Astrophysics Data System, \href
  {https://ui.adsabs.harvard.edu/abs/1976IAUS...73...75P} {73, 75}

\bibitem[\protect\citeauthoryear{Pavlovskii, Ivanova, Belczynski  \&
  Van}{Pavlovskii et~al.}{2016}]{Pavlovskii_2017}
Pavlovskii K.,  Ivanova N.,  Belczynski K.,   Van K.~X.,  2016, \mn@doi
  [Monthly Notices of the Royal Astronomical Society] {10.1093/mnras/stw2786},
  465, 2092

\bibitem[\protect\citeauthoryear{{Paxton}, {Bildsten}, {Dotter}, {Herwig},
  {Lesaffre}  \& {Timmes}}{{Paxton} et~al.}{2011}]{Paxton2011}
{Paxton} B.,  {Bildsten} L.,  {Dotter} A.,  {Herwig} F.,  {Lesaffre} P.,
  {Timmes} F.,  2011, \mn@doi [\apjs] {10.1088/0067-0049/192/1/3}, \href
  {https://ui.adsabs.harvard.edu/abs/2011ApJS..192....3P} {192, 3}

\bibitem[\protect\citeauthoryear{Paxton et~al.,}{Paxton
  et~al.}{2013}]{Paxton2013}
Paxton B.,  et~al., 2013, \mn@doi [The Astrophysical Journal Supplement Series]
  {10.1088/0067-0049/208/1/4}, 208, 4

\bibitem[\protect\citeauthoryear{{Paxton} et~al.,}{{Paxton}
  et~al.}{2015}]{Paxton2015}
{Paxton} B.,  et~al., 2015, \mn@doi [\apjs] {10.1088/0067-0049/220/1/15}, \href
  {https://ui.adsabs.harvard.edu/abs/2015ApJS..220...15P} {220, 15}

\bibitem[\protect\citeauthoryear{{Paxton} et~al.,}{{Paxton}
  et~al.}{2018}]{Paxton2018}
{Paxton} B.,  et~al., 2018, \mn@doi [\apjs] {10.3847/1538-4365/aaa5a8}, \href
  {https://ui.adsabs.harvard.edu/abs/2018ApJS..234...34P} {234, 34}

\bibitem[\protect\citeauthoryear{{Paxton} et~al.,}{{Paxton}
  et~al.}{2019}]{Paxton2019}
{Paxton} B.,  et~al., 2019, \mn@doi [\apjs] {10.3847/1538-4365/ab2241}, \href
  {https://ui.adsabs.harvard.edu/abs/2019ApJS..243...10P} {243, 10}

\bibitem[\protect\citeauthoryear{Pols, Schröder, Hurley, Tout  \&
  Eggleton}{Pols et~al.}{1998}]{Pols_1998}
Pols O.~R.,  Schröder K.-P.,  Hurley J.~R.,  Tout C.~A.,   Eggleton P.~P.,
  1998, \mn@doi [Monthly Notices of the Royal Astronomical Society]
  {10.1046/j.1365-8711.1998.01658.x}, 298, 525

\bibitem[\protect\citeauthoryear{{Sabhahit}, {Vink}, {Higgins}  \&
  {Sander}}{{Sabhahit} et~al.}{2022}]{Sabhahit_2022}
{Sabhahit} G.~N.,  {Vink} J.~S.,  {Higgins} E.~R.,   {Sander} A. A.~C.,  2022,
  \mn@doi [\mnras] {10.1093/mnras/stac1410}, \href
  {https://ui.adsabs.harvard.edu/abs/2022MNRAS.514.3736S} {514, 3736}

\bibitem[\protect\citeauthoryear{{Sana} et~al.,}{{Sana}
  et~al.}{2012}]{Sana2012}
{Sana} H.,  et~al., 2012, \mn@doi [Science] {10.1126/science.1223344}, \href
  {https://ui.adsabs.harvard.edu/abs/2012Sci...337..444S} {337, 444}

\bibitem[\protect\citeauthoryear{Schaerer, Koter  \& Schmutz}{Schaerer
  et~al.}{1995}]{Schaerer_1995}
Schaerer D.,  Koter A.~D.,   Schmutz W.,  1995, Complete Stellar Models for
  Massive Stars.
Springer Netherlands, Dordrecht, pp 300--304,
  \mn@doi{10.1007/978-94-011-0205-6_68}, \url
  {https://doi.org/10.1007/978-94-011-0205-6_68}

\bibitem[\protect\citeauthoryear{{Schootemeijer}, {Langer}, {Grin}  \&
  {Wang}}{{Schootemeijer} et~al.}{2019}]{Schootemeijer_2019}
{Schootemeijer} A.,  {Langer} N.,  {Grin} N.~J.,   {Wang} C.,  2019, \mn@doi
  [\aap] {10.1051/0004-6361/201935046}, \href
  {https://ui.adsabs.harvard.edu/abs/2019A&A...625A.132S} {625, A132}

\bibitem[\protect\citeauthoryear{Scott, Hirschi, Georgy, Arnett, Meakin,
  Kaiser, Ekström  \& Yusof}{Scott et~al.}{2021}]{Scott_2021}
Scott L. J.~A.,  Hirschi R.,  Georgy C.,  Arnett W.~D.,  Meakin C.,  Kaiser
  E.~A.,  Ekström S.,   Yusof N.,  2021, \mn@doi [Monthly Notices of the Royal
  Astronomical Society] {10.1093/mnras/stab752}, 503, 4208

\bibitem[\protect\citeauthoryear{Secunda et~al.,}{Secunda
  et~al.}{2020}]{Secunda_2020}
Secunda A.,  et~al., 2020, \mn@doi [The Astrophysical Journal]
  {10.3847/1538-4357/abbc1d}, 903, 133

\bibitem[\protect\citeauthoryear{{Shenar} et~al.,}{{Shenar}
  et~al.}{2022}]{Shenar_2022}
{Shenar} T.,  et~al., 2022, \mn@doi [\aap] {10.1051/0004-6361/202244245}, \href
  {https://ui.adsabs.harvard.edu/abs/2022A&A...665A.148S} {665, A148}

\bibitem[\protect\citeauthoryear{Smith, Vink  \& de Koter}{Smith
  et~al.}{2004}]{Smith_2004}
Smith N.,  Vink J.~S.,   de Koter A.,  2004, \mn@doi [The Astrophysical
  Journal] {10.1086/424030}, 615, 475

\bibitem[\protect\citeauthoryear{{Spitzer}}{{Spitzer}}{1969}]{Spitzer_1969}
{Spitzer} Lyman J.,  1969, \mn@doi [\apjl] {10.1086/180451}, \href
  {https://ui.adsabs.harvard.edu/abs/1969ApJ...158L.139S} {158, L139}

\bibitem[\protect\citeauthoryear{{Stevenson}, {Sampson}, {Powell},
  {Vigna-G{\'o}mez}, {Neijssel}, {Sz{\'e}csi}  \& {Mandel}}{{Stevenson}
  et~al.}{2019}]{Stevenson_2019}
{Stevenson} S.,  {Sampson} M.,  {Powell} J.,  {Vigna-G{\'o}mez} A.,  {Neijssel}
  C.~J.,  {Sz{\'e}csi} D.,   {Mandel} I.,  2019, \mn@doi [\apj]
  {10.3847/1538-4357/ab3981}, \href
  {https://ui.adsabs.harvard.edu/abs/2019ApJ...882..121S} {882, 121}

\bibitem[\protect\citeauthoryear{Stone, Metzger  \& Haiman}{Stone
  et~al.}{2016}]{Stone_2016}
Stone N.~C.,  Metzger B.~D.,   Haiman Z.,  2016, \mn@doi [Monthly Notices of
  the Royal Astronomical Society] {10.1093/mnras/stw2260}, 464, 946

\bibitem[\protect\citeauthoryear{{Sz\'ecsi}, {Agrawal, Poojan}, {W\"unsch,
  Richard}  \& {Langer, Norbert}}{{Sz\'ecsi} et~al.}{2022}]{szecsi2021}
{Sz\'ecsi} {Agrawal, Poojan} {W\"unsch, Richard}  {Langer, Norbert} 2022,
  \mn@doi [A\&A] {10.1051/0004-6361/202141536}, 658, A125

\bibitem[\protect\citeauthoryear{Tagawa, Haiman  \& Kocsis}{Tagawa
  et~al.}{2020}]{Tagawa_2020}
Tagawa H.,  Haiman Z.,   Kocsis B.,  2020, \mn@doi [The Astrophysical Journal]
  {10.3847/1538-4357/ab9b8c}, 898, 25

\bibitem[\protect\citeauthoryear{Tagawa, Kocsis, Haiman, Bartos, Omukai  \&
  Samsing}{Tagawa et~al.}{2021}]{Tagawa_2021}
Tagawa H.,  Kocsis B.,  Haiman Z.,  Bartos I.,  Omukai K.,   Samsing J.,  2021,
  \mn@doi [The Astrophysical Journal] {10.3847/1538-4357/abd555}, 908, 194

\bibitem[\protect\citeauthoryear{{Vink}, {de Koter}  \& {Lamers}}{{Vink}
  et~al.}{2001}]{Vink2001}
{Vink} J.~S.,  {de Koter} A.,   {Lamers} H.~J.~G.~L.~M.,  2001, \mn@doi [\aap]
  {10.1051/0004-6361:20010127}, \href
  {http://adsabs.harvard.edu/abs/2001A%26A...369..574V} {369, 574}

\bibitem[\protect\citeauthoryear{{Vink}, {Higgins}, {Sander}  \&
  {Sabhahit}}{{Vink} et~al.}{2021}]{Vink_2021}
{Vink} J.~S.,  {Higgins} E.~R.,  {Sander} A. A.~C.,   {Sabhahit} G.~N.,  2021,
  \mn@doi [\mnras] {10.1093/mnras/stab842}, \href
  {https://ui.adsabs.harvard.edu/abs/2021MNRAS.504..146V} {504, 146}

\bibitem[\protect\citeauthoryear{{Webbink}}{{Webbink}}{1984}]{Webbink_1984}
{Webbink} R.~F.,  1984, \mn@doi [\apj] {10.1086/161701}, \href
  {https://ui.adsabs.harvard.edu/abs/1984ApJ...277..355W} {277, 355}

\bibitem[\protect\citeauthoryear{{Wong}, {Breivik}, {Kremer}  \&
  {Callister}}{{Wong} et~al.}{2021}]{Wong_2021}
{Wong} K. W.~K.,  {Breivik} K.,  {Kremer} K.,   {Callister} T.,  2021, \mn@doi
  [\prd] {10.1103/PhysRevD.103.083021}, \href
  {https://ui.adsabs.harvard.edu/abs/2021PhRvD.103h3021W} {103, 083021}

\bibitem[\protect\citeauthoryear{{Woosley}}{{Woosley}}{2017}]{Woosley_2017}
{Woosley} S.~E.,  2017, \mn@doi [\apj] {10.3847/1538-4357/836/2/244}, \href
  {https://ui.adsabs.harvard.edu/abs/2017ApJ...836..244W} {836, 244}

\bibitem[\protect\citeauthoryear{Xin, Renzo  \& Metzger}{Xin
  et~al.}{2022}]{Xin_2022}
Xin C.,  Renzo M.,   Metzger B.~D.,  2022, \mn@doi [Monthly Notices of the
  Royal Astronomical Society] {10.1093/mnras/stac2551}, 516, 5816

\bibitem[\protect\citeauthoryear{Xu \& Li}{Xu \& Li}{2010}]{Xu_2010}
Xu X.-J.,  Li X.-D.,  2010, \mn@doi [The Astrophysical Journal]
  {10.1088/0004-637x/716/1/114}, 716, 114

\bibitem[\protect\citeauthoryear{Yang et~al.,}{Yang et~al.}{2019}]{Yang_2019}
Yang Y.,  et~al., 2019, \mn@doi [Phys. Rev. Lett.]
  {10.1103/PhysRevLett.123.181101}, 123, 181101

\bibitem[\protect\citeauthoryear{{Zevin} et~al.,}{{Zevin}
  et~al.}{2021}]{Zevin_2021}
{Zevin} M.,  et~al., 2021, \mn@doi [\apj] {10.3847/1538-4357/abe40e}, \href
  {https://ui.adsabs.harvard.edu/abs/2021ApJ...910..152Z} {910, 152}

\bibitem[\protect\citeauthoryear{{Zi{\'o}{\l}kowski}}{{Zi{\'o}{\l}kowski}}{2008}]{Ziolkowski_2008}
{Zi{\'o}{\l}kowski} J.,  2008, \mn@doi [Chinese Journal of Astronomy and
  Astrophysics Supplement] {10.48550/arXiv.0808.0435}, \href
  {https://ui.adsabs.harvard.edu/abs/2008ChJAS...8..273Z} {8, 273}

\bibitem[\protect\citeauthoryear{{de Jager}, {Nieuwenhuijzen}  \& {van der
  Hucht}}{{de Jager} et~al.}{1988}]{deJager1988}
{de Jager} C.,  {Nieuwenhuijzen} H.,   {van der Hucht} K.~A.,  1988, \aaps,
  \href {https://ui.adsabs.harvard.edu/abs/1988A&AS...72..259D} {72, 259}

\bibitem[\protect\citeauthoryear{de Mink \& Belczynski}{de~Mink \&
  Belczynski}{2015}]{Mink_2015}
de Mink S.~E.,  Belczynski K.,  2015, \mn@doi [The Astrophysical Journal]
  {10.1088/0004-637x/814/1/58}, 814, 58

\bibitem[\protect\citeauthoryear{de Mink \& Mandel}{de~Mink \&
  Mandel}{2016}]{deMinkMandel_2016}
de Mink S.~E.,  Mandel I.,  2016, \mn@doi [Monthly Notices of the Royal
  Astronomical Society] {10.1093/mnras/stw1219}, 460, 3545

\bibitem[\protect\citeauthoryear{{van Son} et~al.,}{{van Son}
  et~al.}{2022}]{vanSon_2022}
{van Son} L.~A.~C.,  et~al., 2022, \mn@doi [\apj] {10.3847/1538-4357/ac64a3},
  \href {https://ui.adsabs.harvard.edu/abs/2022ApJ...931...17V} {931, 17}

\makeatother
\end{thebibliography}




\appendix

\section{Interaction radius distribution for BH-BH merger progenitors}
\label{App:IntRad}

\begin{figure*}
	\includegraphics[width=18 cm]{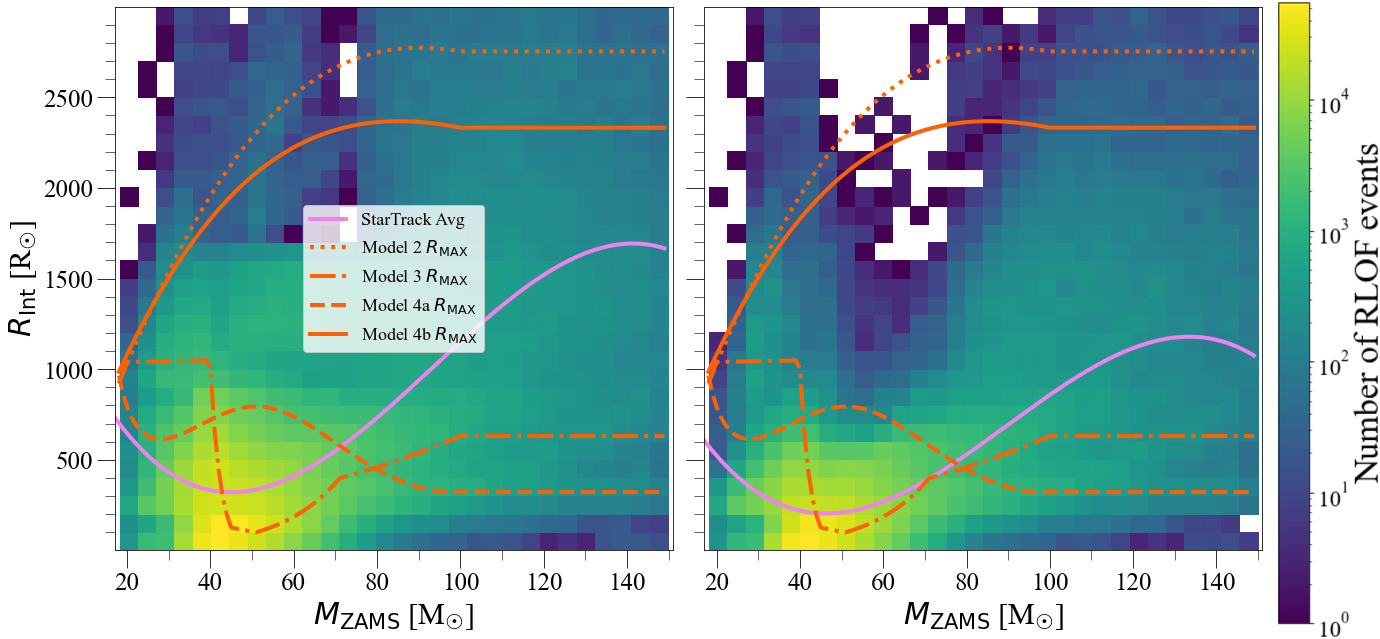}
    \caption{2D histogram showing the distribution of all RLOF events (l.h.s. plot) and only stable RLOF events (r.h.s. plot) as a function of $R_{\text{Int}}$ and $M_{\text{ZAMS}}$. Both plots were rescaled to include only the events with $R_{\text{Int}}$ < 3000\,R$_\odot$. The orange lines show $R_{\text{MAX}}$ as a function of $M_{\text{ZAMS}}$ for our Model 2, Model 3, Model 4a and Model 4b, while the pink line represents the average $R_{\text{Int}}$ values as a function of $M_{\text{ZAMS}}$ (Eq. \ref{eq:Rint_ZAMS_RLOFALL} and \ref{eq:Rint_ZAMS_RLOFstab}).
    }
    \label{fig:RLOF}
\end{figure*}

The {\tt StarTrack} average $R_{\text{Int}}$ for all types of RLOF events, as shown by the pink line in the left plot from Figure \ref{fig:RLOF}, follows the following relation:

\begin{equation}
    R_{\text{Int}} = -3.1M_{\rm ZAMS}^3\times10^{-3} + 0.8M_{\text{ZAMS}}^2 - 58.4M_{\text{ZAMS}} - 1495.6
    \label{eq:Rint_ZAMS_RLOFALL}
\end{equation}

The average for only stable RLOF events, as shown by the pink line in the left plot from Figure \ref{fig:RLOF}, is described by the following equation:

\begin{equation}
    R_{\text{Int}} = -2.9M_{\rm ZAMS}^3\times10^{-3} + 0.8M_{\text{ZAMS}}^2 - 54.4M_{\text{ZAMS}} - 1317.7
    \label{eq:Rint_ZAMS_RLOFstab}
\end{equation}

\begin{figure*}
    \centering
	    \includegraphics[width=11.7 cm]{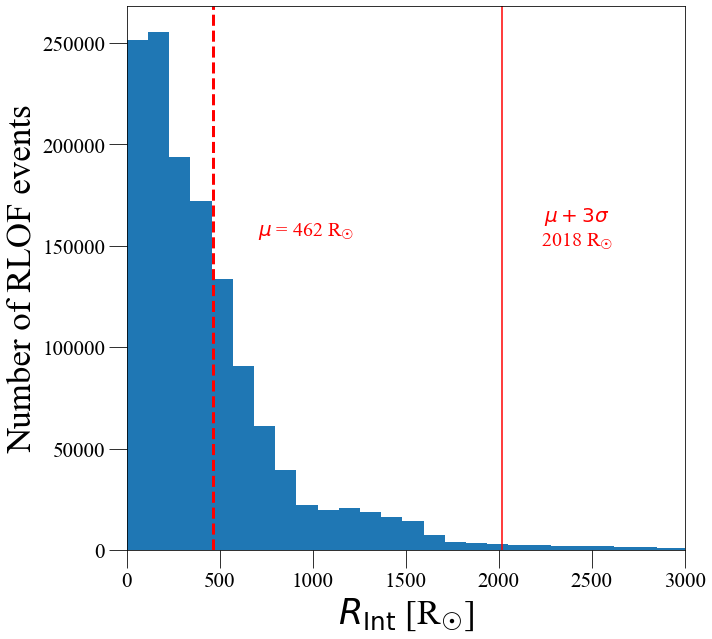}
        \caption{Distribution of the interaction radii for all RLOF events from our standard Model 1 simulations. The dashed vertical red line represents the median value at $\sim462$\,R$_\odot$, while the continuous vertical line represents a range of three standard deviations from the median.}
        \label{fig:AllMT}
\end{figure*}

\begin{figure*}
    \centering
	    \includegraphics[width=11.7 cm]{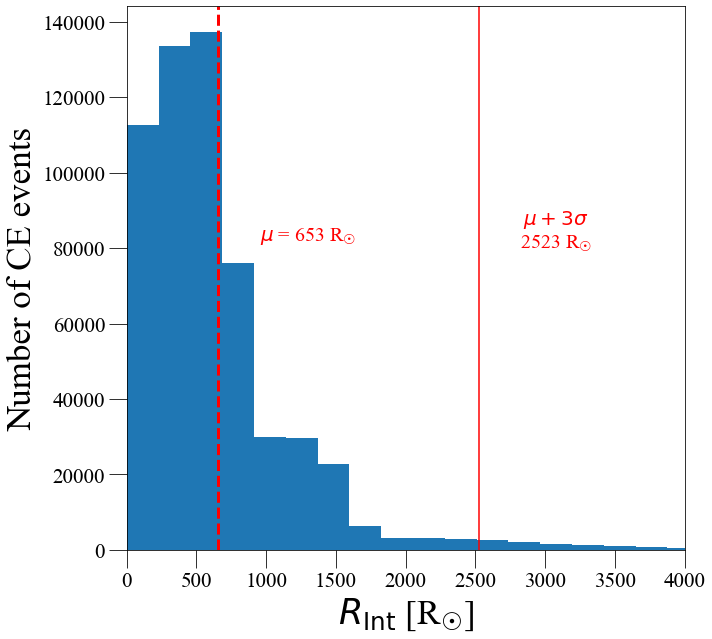}
        \caption{Distribution of the interaction radii for CE events from our standard Model 1 simulations. The dashed vertical red line represents the median value at $\sim653$\,R$_\odot$, while the continuous vertical line represents a range of three standard deviations from the median.}
        \label{fig:CE}
\end{figure*}

\begin{figure*}
    \centering
	    \includegraphics[width=11.7 cm]{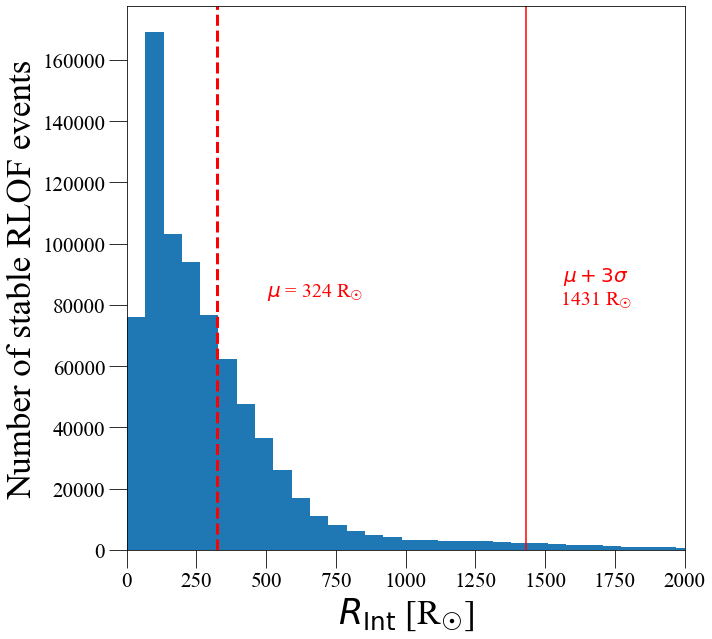}
        \caption{Distribution of the interaction radii for stable RLOF events from our standard Model 1 simulations. The dashed vertical red line represents the median value at $\sim324$\,R$_\odot$, while the continuous vertical line represents a range of three standard deviations from the median.}
        \label{fig:StableRLOF}
\end{figure*}

\bsp	
\label{lastpage}
\end{document}